\documentclass[final,superscriptaddress,english,twocolumn,amssymb,aps,prb,
longbibliography]{revtex4-2}
\usepackage{graphicx}
\usepackage{natbib}
\usepackage{amsmath}
\usepackage{amssymb}
\usepackage{appendix}
\usepackage{bm}
\usepackage{overpic}
\usepackage[dvipsnames]{xcolor}{\huge }
\usepackage[section]{placeins}
\definecolor{darkblue}{rgb}{0,0,.65}
\definecolor{darkgreen}{rgb}{0.28,0.41,0.19}

\usepackage{multirow}

\usepackage{nicefrac}
\usepackage[%
    pdfauthor={Imre Hagymasi},%
  pdfstartview=FitH,%
  breaklinks=true,%
  bookmarks=true,%
  colorlinks=true,%
  anchorcolor=black,%
  citecolor=blue,
  filecolor=black,%
  menucolor=black,%
  urlcolor=darkblue,%
  linkcolor=blue,%
 ]{hyperref}
\usepackage[all]{hypcap} %

\usepackage[capitalize]{cleveref}

\newcommand{\GammaL}{\Gamma^{\textrm{(L)}}}
\newcommand{\GammaR}{\Gamma^{\textrm{(R)}}}

\graphicspath{{images/}}
\begin{document}

\title{Phase diagram of the antiferromagnetic $J_1$-$J_2$ spin-$1$ pyrochlore Heisenberg model}

\author{Imre Hagym\'asi}
\affiliation{Helmholtz-Zentrum Berlin f\"ur Materialien und Energie, Hahn-Meitner Platz 1, 14109 Berlin, Germany}
\affiliation{Dahlem Center for Complex Quantum Systems and Fachbereich Physik, Freie Universit\"at Berlin, Arnimallee 14, 14195 Berlin, Germany}
\affiliation{Institute 
for Solid State
Physics and Optics, Wigner Research Centre for Physics, Budapest H-1525 P.O. 
Box 49, Hungary
}
\author{Nils Niggemann}
\affiliation{Helmholtz-Zentrum Berlin f\"ur Materialien und Energie, Hahn-Meitner Platz 1, 14109 Berlin, Germany}
\affiliation{Dahlem Center for Complex Quantum Systems and Fachbereich Physik, Freie Universit\"at Berlin, Arnimallee 14, 14195 Berlin, Germany}
\affiliation{Department of Physics and Quantum Centers in Diamond and Emerging Materials (QuCenDiEM) group, Indian Institute of Technology Madras, Chennai 600036, India}
\author{Johannes Reuther}
\affiliation{Helmholtz-Zentrum Berlin f\"ur Materialien und Energie, Hahn-Meitner Platz 1, 14109 Berlin, Germany}
\affiliation{Dahlem Center for Complex Quantum Systems and Fachbereich Physik, Freie Universit\"at Berlin, Arnimallee 14, 14195 Berlin, Germany}

\date{\today}

\begin{abstract}
We study the phase diagram of the antiferromagnetic $J_1$-$J_2$ Heisenberg model on the pyrochlore lattice with $S=1$ spins at zero and finite temperatures. We use a combination of complementary state-of-the-art quantum many-body approaches such as density matrix renormalization group (DMRG), density-matrix purification and pseudo-Majorana functional renormalization group (PMFRG). We present an efficient approach to preserve the applicability of the PMFRG for spin-1 systems at finite temperatures despite the inevitable presence of unphysical spin states. The good performance of our methods is first demonstrated for the nearest-neighbor pyrochlore Heisenberg model where the finite temperature behavior of the specific heat and uniform susceptibility show excellent agreement within PMFRG and density-matrix purification. Including an antiferromagnetic second neighbor coupling we find that the non-magnetic ground-state phase of the nearest neighbor model extents up to $J_2/J_1 \sim 0.02 $ within DMRG, beyond which magnetic ${\bm k}=0$ long-range order sets in. Our PMFRG calculations find the phase transition in a similar regime $J_2/J_1\sim 0.035(8)$ which, together with the DMRG result, provides a strong argument for the existence of a small but finite non-magnetic ground-state phase in the spin-1 pyrochlore Heisenberg model. We also discuss the origin of discrepancies between different versions of the functional renormalization group concerning the location of this phase transition.
\end{abstract}
\maketitle

\section{Introduction}
The understanding of magnetically frustrated quantum systems is still among the central problems of condensed matter physics. Due to frustration, the ground state of a quantum spin system is not simply related to a classical long-range ordered state that simultaneously minimizes the local energies on all lattice bonds, like in a conventional antiferromagnet. Instead frustration induces a complex global interplay of spins which may results in a highly entangled ground state such as a quantum spin liquid~\cite{balents2010_spinliquid}.

The most prominent systems where frustration emerges from the geometrical structure are the Heisenberg models on the kagom\'e lattice (in two dimensions) and on the pyrochlore lattice (in three dimensions) with antiferromagnetic couplings between neighboring spins. Tremendous effort has been put into understanding these systems both at the classical and quantum level \cite{jiang_prl_2008,pollmann_prx_2017,yan_science_2011,schollwock_prl_2012,lauchli_kagome_2019,kim_prb_2008,sobral_1998,canals_lacroix_prb_2000,candra_numerics_pyrochlore_correlations_2018,iqbal_quantum_2019,muller_thermodynamics_2019,tsunetsugu_theory_2017,ross_quantum_2011,Rau_Gingras_review_2019,derzhko_adapting_2020}. In particular, the Ising model on the pyrochlore lattice realizes the celebrated spin ice state \cite{harris_geometrical_1997}, a classical spin liquid which hosts fractional excitations and an emergent U(1) gauge field \cite{castelnovo_spin_2012}.

In the extreme spin-1/2 quantum limit, recent density-matrix renormalization group (DMRG) \cite{hagymasi_prl_2021}, variational Monte Carlo \cite{astrakhantsev_broken-symmetry_2021} and pseudo-fermion functional renormalization group (PFFRG) \cite{hering2021dimerization} calculations show strong indications that the ground state of the pyrochlore Heisenberg antiferromagnet with only nearest-neighbor interactions ($J_1$) is non-magnetic but breaks the $C_3$ and/or inversion symmetries of the lattice. These findings provide a compelling argument against a spin liquid scenario. Very recently, this has been further substantiated by the proposal of a new family of valence bond crystals which produce competitive energies with the aforementioned numerical calculations \cite{schaefer2022abundance}. Even though all these studies cannot finally answer the question about the system's ground state, they significantly limit the possible scenarios and highlight the competition between different states.

An interesting extension is the inclusion of second neighbor Heisenberg couplings ($J_2$) on the pyrochlore lattice. These interactions enrich the competition between different states even further as they increase the tendency towards magnetic long-range order. Specifically, in the classical limit even an infinitesimally small antiferromagnetic $J_2$ makes the system enter the $\bm{k}=0$ magnetic phase \cite{iqbal_quantum_2019}, where all tetrahedra related by lattice translations exhibit the same spin configuration. In the opposite limit of $S=1/2$ quantum spins the non-magnetic phase was found to survive in a remarkably large region $J_2/J_1 \leq 0.22(3) $ according to PFFRG calculations~\cite{iqbal_quantum_2019}. Later, variational Monte Carlo studies revealed a considerably smaller extent of the magnetically disordered regime, occurring only for $J_1/J_2\leq 0.0295(30)$~\cite{astrakhantsev_broken-symmetry_2021}. Despite the numerous indications of an extended non-magnetic regime in the $S=1/2$ model, already at $S=1$ it is {\it a priori} unclear whether a magnetically disordered phase can survive, since increasing the spin magnitude decreases quantum fluctuations. Nevertheless, within PFFRG an extended non-magnetic phase was found in the $S=1$ case where the system did not show magnetic order up to $J_2/J_1 = 0.09(2)$~\cite{iqbal_quantum_2019}.

Our goal in this paper is to examine the effect of second neighbor couplings $J_2$ in a $S=1$ pyrochlore Heisenberg model in more detail, both at zero and finite temperatures making use of recent methodological advances. The $S=1$ case is particularly relevant to experiments: The compound NaCaNi$_2$F$_7$ realizes an (almost) isotropic $S=1$ pyrochlore Heisenberg model and does not exhibit magnetic ordering down to $100$ mK \cite{plumb_continuum_2019}.
We apply two very different approaches in our numerical studies, the pseudo-Majorana functional renormalization group (PMFRG) technique~\cite{niggemann2021majorana,niggemann2021quantitative} and matrix-product-state methods (density-matrix purification and DMRG) \cite{white_1992,white_1993,white_purification_2005,cirac_purification_2004,schollwock_review_2005,schollwock_review_2011,hallberg_review}. It has been shown that the PMFRG approach overcomes the obstacle of the more standard (complex fermionic) PFFRG, since the employed representation of spin operators in terms of Majorana fermions does not introduce unphysical states for the $S=1/2$ case and produces accurate results at finite temperatures \cite{Niggemann2021a,niggemann2021quantitative}. However, since an exact bilinear Majorana representation for $S=1$ spins does not exist~\cite{schaden2023bilinear}, the formalism used here cannot avoid unphysical states. We demonstrate that the PMFRG can nevertheless be faithfully applied to spin-1 systems by shifting the unphysical states to higher energies such that they do not affect finite-temperature results.

As a first benchmark, we address the finite-temperature properties of the nearest-neighbor pyrochlore Heisenberg model and find excellent agreement for the specific heat and susceptibility within the two methods. In the presence of the antiferromagnetic $J_2$ interaction PMFRG predicts the nonmagnetic phase to be stable up to $J_2/J_1\sim0.035(8)$ before $\bm{k}=0$ order sets in. This is about one third of the previous estimate $J_2/J_1\sim0.09(2)$ from PFFRG \cite{iqbal_pyrochlore_s05_2017}. We argue that the value from PMFRG can be considered more accurate than the PFFRG estimate. The quality of the PMFRG result is further underpinned by DMRG which locates the phase transition in a similar region $J_2/J_1\sim0.02$. Taken together, the independent yet consistent identification of the phase transition by two fundamentally different approaches provides sound evidence for the existence of a small but finite non-magnetic ground state phase in the $S=1$ pyrochlore Heisenberg model. In general, our studies demonstrate the capabilities of state-of-the-art numerical methods in accurately treating even complex interacting systems such as frustrated quantum spins on complicated three-dimensional networks.

The paper is organized as follows. In Sec.~\ref{sec:model} we overview the basics of the pyrochlore Heisenberg model and the methods applied in this paper. Section~\ref{sec:results} presents the results of our study where Sec.~\ref{sec:finite_T} considers thermodynamic properties of the $J_1$-only model while Sec.~\ref{sec:j2} discusses the effects of the $J_2$ coupling on the system's ground state properties. In Sec.~\ref{sec:comparison} we compare and interpret our results in the light of previous PFFRG studies. Finally, in Sec.~\ref{sec:conclusion} our conclusions are presented.
\begin{figure}[t]
    \centering
    \includegraphics[width=0.6\columnwidth]{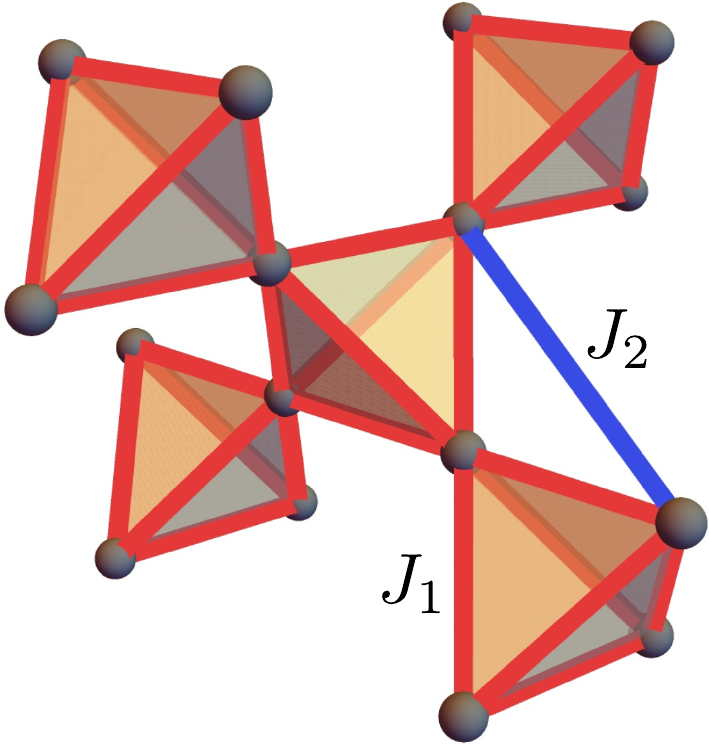}
    \caption{Pyrochlore lattice with nearest neighbor bonds $J_1$ shown in red and one representative second neighbor bond $J_2$ in blue.}
        \label{fig:couplings}
\end{figure}

\section{Model and methods}\label{sec:model}
\subsection{Pyrochlore Heisenberg model}
We consider the following Hamiltonian on the pyrochlore lattice,
\begin{equation}\label{eq_ham}
	H = J_1 \sum_{\langle i, j\rangle} \bm{S}_i \cdot \bm{S}_j + J_2 \sum_{\langle\langle i, j\rangle\rangle} \bm{S}_i \cdot \bm{S}_j, 
\end{equation}
where $\bm{S}_i$ is a spin-1 operator on site $i$ and we use natural units, $J_1=1$, $k_B=1$ and $\hbar=1$. Here, $J_1$ denotes the coupling on nearest neighbor pyrochlore bonds $\langle i,j\rangle$  and $J_2$ is the coupling on second neighbor pyrochlore bonds $\langle\langle i,j\rangle\rangle$ as shown in Fig.~\ref{fig:couplings}. The pyrochlore lattice is spanned by an fcc lattice with a tetrahedral crystal basis defined by the vectors $\bm{b}_0=0$, $\bm{b}_i = \frac{1}{2} \bm{a}_i$ where the fcc lattice vectors are given by $\bm{a}_1 = \frac{1}{2} (1,1,0)^T$, $\bm{a}_2 = \frac{1}{2} (1,0,1)^T$, $\bm{a}_3 = \frac{1}{2} (0,1,1)^T$. Thus every lattice point can be expressed as $\bm{R}_i\equiv\bm{R}_{l,n_1,n_2,n_3} = n_1 \bm{a}_1 +  n_2 \bm{a}_2 +  n_3 
\bm{a}_3 + \bm{b}_l,$
with integer $n_1,n_2,n_3$ and $l \in \{0,1,2,3\}$.

\subsection{DMRG and density-matrix purification}
Firstly, to study finite temperatures we use the density-matrix purification technique. This approach relies on the fact that the density matrix can be interpreted as a partial trace of the Schmidt decomposition of an enlarged system, which is itself in a pure state \cite{white_purification_2005,schollwock_review_2011}. This enlargement is done by adding auxiliary sites (ancillas) to the physical system, which need to be traced out to obtain the density matrix. In this language, the infinite temperature state can be written as a maximally entangled state between each site and its own ancilla. We obtain the density matrix at a given temperature by performing imaginary time evolution on the physical sites of the enlarged system and tracing out the ancilla sites. To carry out the imaginary time evolution we use the time-dependent variational principle (TDVP) \cite{haegeman_2011,haegeman_2016} with the two-site update and follow the steps described in Ref.~\cite{schafer_pyrochlore_2020} to ensure that the TDVP does not produce false results~\cite{hubig_review_2019}.

Secondly, we use the DMRG method \cite{white_1992,white_1993,schollwock_review_2005,schollwock_review_2011,hallberg_review} to study the ground-state properties. This method relies on the assumption that the wave function can be faithfully represented by a matrix-product state form. The algorithm optimizes the energy by local updates and becomes exact if we allow that the matrices can be arbitrarily large. We use the two-site DMRG as well as the single-site variant with the subspace expansion to optimize the wave function \cite{hubig_2015,hubig17:_symmet_protec_tensor_networ}.

For both approaches we exploit the SU(2) symmetry of the Hamiltonian, which enables a better compression of the wave function. During the imaginary time evolution we keep up to 10000 SU(2) states, while in our DMRG simulations this is increased to 16000. We consider a 32-site and a 48-site cluster with periodic boundary conditions where the former respects all the point group symmetries of the fcc lattice. These clusters are mapped to a one-dimensional topology with long-range interactions so that the DMRG method can be applied. The geometry and mapping of the clusters are identical to those in Ref.~\cite{hagymasi2022prb}. To extrapolate the energies towards the infinite bond dimension limit we use the two-site variance as an error measure \cite{hubig_prb_2018}. For finite temperatures the extrapolation of the specific heat and susceptibility is performed as a function of the inverse bond dimension. In the purification approach, we calculate the heat capacity (per site), $C_V$ from the derivative of the energy with respect to temperature,
  \begin{equation}
      C_V = \frac{1}{N}\frac{\partial \langle H \rangle_\beta}{\partial T}=-\beta^2\frac{1}{N}\frac{\partial \langle H \rangle_\beta}{\partial \beta}, 
      \label{eq:spec_heat}
  \end{equation}
where $N$ is the total number of sites, $\beta$ is the inverse temperature and $\langle \dots \rangle_{\beta}$ means the average with respect to the density-matrix at inverse temperature $\beta$. In practice, the heat capacity $C_V$ is obtained by using the rightmost expression in this equation and taking the derivative of the cubic spline interpolation of the energy as a function of the inverse temperature. Furthermore, the equal-time spin structure factor $S(\bm{q})$ is calculated via
\begin{equation}
    S(\bm{q})= \frac{1}{N} \sum_{i j} \langle \bm{S}_i\cdot 
\bm{S}_j\rangle_\beta \cos\left[\bm{q}\cdot 
\left(\bm{R}_i-\bm{R}_j\right)\right],\label{eq:ssf_equal_time}
\end{equation}
and the uniform susceptibility (per site), $\chi$, is defined by 
  \begin{equation}
 \chi=\frac{\beta}{3N}\sum_{ij}\langle \bm{S}_i \cdot \bm{S}_j\rangle_\beta=\frac{\beta}{3}S({\bm q}=0 ).
\end{equation} 

\subsection{PMFRG}
As a complementary approach that is less affected by finite-size effects, we employ the PMFRG. This method is well suited for three dimensional frustrated magnets for which it has proven both flexible and quantitatively reliable, and in particular allows for an unbiased detection of phase transitions via finite-size scaling~\cite{niggemann2021quantitative, schneider2023temperature,niggemann2021majorana,muller2023pseudo}. 
The PMFRG uses a mapping of $S=1/2$ spin operators onto three flavors of Majorana fermions satisfying $\{\eta^\alpha_i,\eta^\beta_j\} = \delta_{ij}\delta^{\alpha \beta}$ (with $\alpha,\beta\in\{x,y,z\}$), such that $S^x_i = -i \eta^y_i\eta^z_i$, and the other spin components follow by cyclic permutations of $x$, $y$, $z$. Unlike other fermionic spin representations, this Majorana rewriting has the advantage that it does not introduce any unphysical states in the Hilbert space which in particular allows for an application at finite temperatures.

Although the PMFRG implements spin-$1/2$ operators exactly, a  spin-$1$ representation cannot be achieved without the introduction of unphysical states~\cite{schaden2023bilinear}. Here, we follow the approach of Ref.~\cite{Baez17} where an effective higher spin $S_\text{eff}$ is implemented by introducing $2S_\text{eff}$ identical replicas $\bm{S}_{\mu}$ of spin-$1/2$ degrees of freedom on each site where $\mu\in\{1,2,\ldots,2S_\text{eff}\}$ is the replica index. According to the rules of addition of angular momenta this generates total spin quantum numbers on each site given by $S \in \{S_\text{eff},S_\text{eff}-1,\dots,0\}$ for integer $S_\text{eff}$.
Unphysical states with $S < S_\text{eff}$ can be excluded via a level repulsion term $\sim A (\sum_\mu {\bm S}_{\mu})^2$ on each site (see the \ref{app:Appendix} 
for details) which for $A<0$ acts as a ferromagnetic coupling between the replica spin-$1/2$ sites and produces an excitation gap $\sim |A|$ to unphysical states. 

In the limit $A \rightarrow -\infty$, unphysical states lie at infinite energy and thus cannot be excited at any finite temperature. However, PMFRG becomes inapplicable in this limit since the involved approximations (truncation of flow equations for the fermionic vertex functions) are valid only at temperatures that are not too small compared to the relevant energy scales of the Hamiltonian $J_1$, $J_2$ and $|A|$. In practice, it is best to choose $A \propto -T$, which ensures that unphysical states remain always suppressed while also keeping the ratio between relevant energy scales of the Hamiltonian and the temperature small. Using this approach, we find that the temperature dependence of the heat capacity and susceptibility seem to converge around $A/T \approx -2.5$ upon varying $A/T$, see the \ref{app:Appendix}. Precisely at this ratio $A/T \approx -2.5$ we also obtain perfect agreement with our DMRG results.

After appropriately adjusting the level repulsion term, the resulting interacting Majorana Hamiltonian is solved via the standard FRG approach, in which an artificial infrared cutoff $\Lambda$ of Matsubara frequencies is introduced in the bare Majorana propagator. Initially, at $\Lambda = \infty$, the fermionic propagation is thus fully removed and the system behaves trivially. To access the system's evolution in $\Lambda$ down to the physical limit $\Lambda=0$, differential flow equations for the fermionic one-particle irreducible vertex functions are numerically solved as a function of $\Lambda$. To obtain a finite set of differential equations, in the standard one-loop truncation~\cite{muller2023pseudo} effective interactions between three or more spins are neglected, although such terms could in principle be generated during the renormalization group flow. An inclusion of certain contributions of such effective three-spin interactions can be achieved by a so-called two-loop truncation which can also be generalized to multiloop schemes~\cite{muller2023pseudo}. From the viewpoint of perturbation theory, a truncation of flow equations is generally justified for temperatures that are not too small compared to the dominant interaction, here given by $J_1$.

From the two-particle vertex, we compute the equal-time spin structure factor $S(\bm q)$ as defined in Eq.~(\ref{eq:ssf_equal_time}).
Other observables, such as the specific heat, are obtained via numerical derivatives of the interacting free energy as detailed in Ref.~\cite{niggemann2021majorana}. 
Within PMFRG, phase transitions between magnetically ordered and paramagnetic phases are typically identified via a critical scaling of the susceptibility, or, alternatively, the correlation length $\xi$ which can be obtained from the equal-time spin structure factor $S(\bm q)$ via
\begin{equation}
    \xi=\frac{L}{2\pi}\sqrt{\frac{S(\bm{q}^\star)}{S(\bm{q}^\star+ \frac{2\pi}{L} {\bm e}_{\bm q^\star})}-1}.\label{eq:correlationRatio}
\end{equation}
Here, $L$ is the maximal distance beyond which spin correlations are set to zero in the simulations and $\bm q^\star$ is the point in reciprocal space where $S(\bm q)$ is maximal which corresponds to the wave vector of the magnetic order to be probed. Furthermore, ${\bm e}_{\bm q^\star}$ is a unit vector in reciprocal space that points towards the direction of deepest descend away from $\bm q^\star$.  Crucially, a finite $L$ has a similar effect than a finite system size~\cite{sandvik_computational_2010} such that precisely at the critical temperature the correlation length behaves as $\xi \sim L$. This criterion is used to detect critical ordering temperatures in numerical runs with varying $T$ and $L$.

\section{Results}\label{sec:results}
\subsection{Finite-temperature properties of the nearest neighbor model}\label{sec:finite_T}
\begin{figure}[t]
    \centering
    \includegraphics[width=\columnwidth]{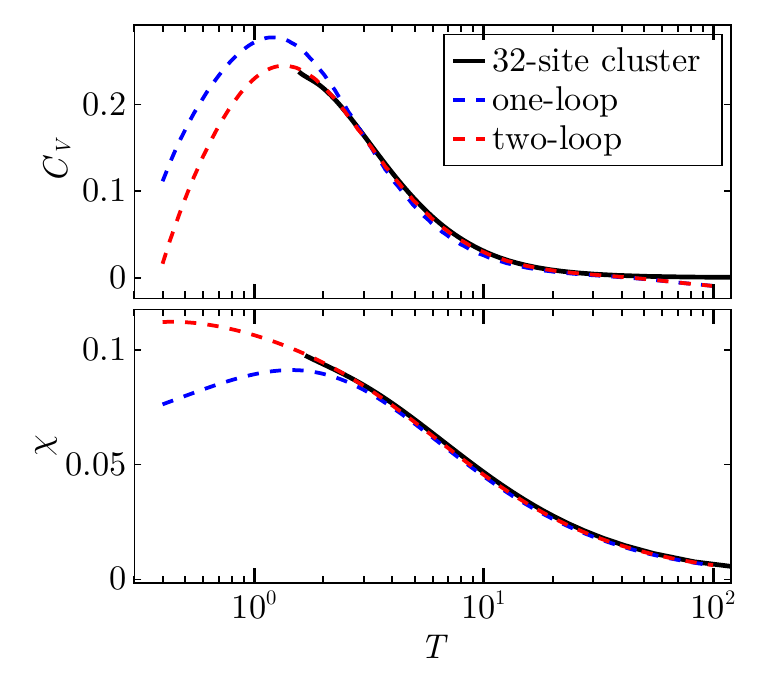}
    \caption{Specific heat $C_V$ (upper panel) and uniform susceptibility $\chi$ (lower panel) for the nearest neighbor pyrochlore Heisenberg model ($J_2=0$) from one-loop and two-loop PMFRG as well as from extrapolated purification for the 32-site cluster.}
        \label{fig:cv_chi}
\end{figure}

In this subsection we address the finite-temperature properties of the nearest neighbor pyrochlore Heisenberg model, i.e., we set $J_2=0$. The ground state of this system was found to be non-magnetic in several previous works~\cite{iqbal_quantum_2019,muller_thermodynamics_2019,hagymasi2022prb}. The investigation here also serves as a benchmark for the PMFRG method to check whether the level-repulsion term can correctly eliminate contributions from the unphysical states in thermodynamic quantities such as the specific heat and the susceptibility. 
In Fig.~\ref{fig:cv_chi} we show the results for these quantities obtained by the purification approach and by PMFRG. 

As we approach $T\sim 2$ a maximum seems to be formed in the heat capacity in agreement with the rotation-invariant Green's function method \cite{muller_thermodynamics_2019}. Since the heat capacity peak occurs at temperatures on the order of the coupling $J_1$, the system turns into a strongly correlated quantum magnet in this temperature region. This is also indicated by the fact that the truncation error in the imaginary time evolution grows rapidly up to $\mathcal{O}(10^{-2})$, which renders the application of the purification technique at lower temperatures unreliable. The PMFRG results are in remarkably good agreement with the ones from the purification approach, especially for the two-loop truncation scheme, which was previously found suitable for the $S=1/2$ nearest neighbor pyrochlore Heisenberg model~\cite{niggemann2021quantitative}. Note that in PMFRG the specific heat is obtained via a numerical second derivative of the free energy, and thus requires a numerically accurate solution of the flow equations, making it prone to inaccuracies from error propagation. Due to such effects we find a slightly negative heat capacity at the highest temperatures $T\sim 100$, where the physical contribution of the interaction correction to the free energy is vanishingly small but the dominant interaction from the level repulsion $A \sim T$ is still considerable. Likewise, at the lowest temperatures, $T \lesssim 0.3$ errors from the truncation of PMFRG flow equations also lead to an unphysical heat capacity $C_V<0$. We have omitted data at such low temperatures in Fig.~\ref{fig:cv_chi}. Nonetheless, the overall agreement between the two very different methods serves as an indication that even for spin magnitudes $S>1/2$, PMFRG is a reliable approach to address finite-temperature properties.

\subsection{Effect of the $J_2$ coupling}\label{sec:j2}
As mentioned before, an antiferromagnetic next-nearest neighbor coupling $J_2$ enhances spin correlations of so-called $\bm{k}=0$ type~\cite{iqbal_quantum_2019}. The corresponding classical $\bm{k}=0$ order has ferromagnetic spin arrangements in each of the four sublattices of the pyrochlore lattice, individually. On the other hand, the relative orientations of the four spins in each tetrahedron fulfill the spin ice rule, which means that the sum of the four spins in each tetrahedron vanishes. The $\bm{k}=0$ order manifests in magnetic Bragg peaks in the spin structure factor at ${\bm q}=(4\pi,0,0)$ and symmetry-equivalent points in the extended Brillouin zone. Our main goal in this subsection is to identify and locate the phase boundary where $J_2$ interactions drive $\bm{k}=0$ magnetic order in the $S=1$ system.

\begin{figure}[t]
    \centering
    \includegraphics[width=\columnwidth]{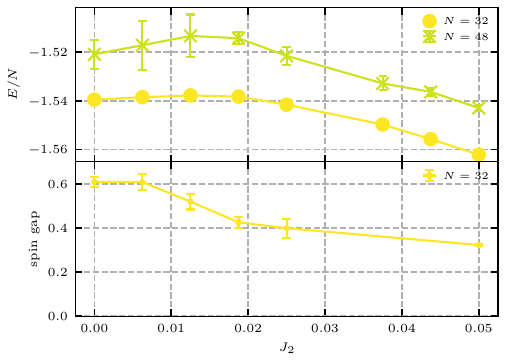}
    \caption{DMRG ground-state energies per site for the 32- and 48-site clusters (upper panel) and spin gap for the 32-site cluster (lower panel). The spin gap is not calculated for the 48-site cluster due to large numerical costs. The error bars are defined as half the distance between the best variational energy and the extrapolated one.}
        \label{fig:energies}
\end{figure}

\begin{figure}[t]   
\begin{overpic}[width=\columnwidth]
{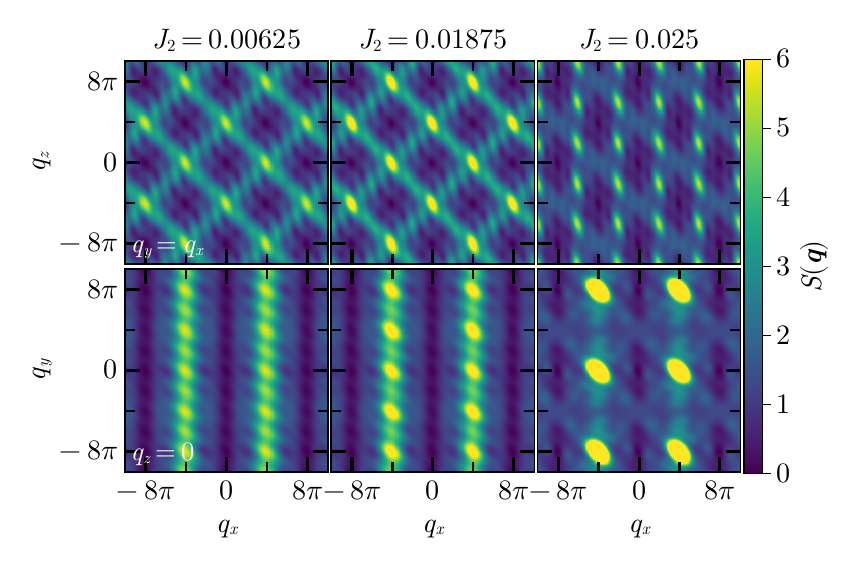}
\put(2,69){(a)}
\end{overpic}\\[0.45cm]    
\begin{overpic}[width=\columnwidth]
{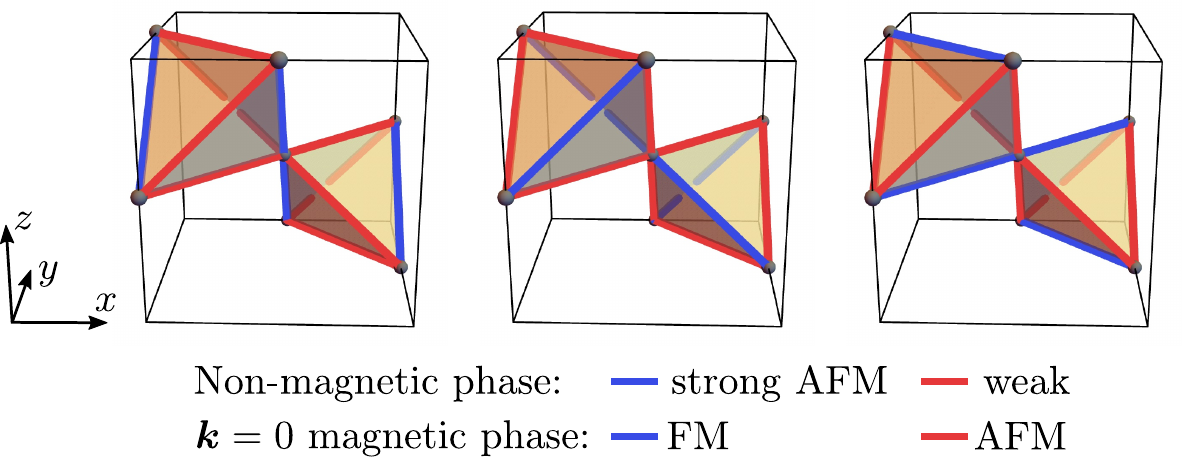}
\put(2,35){(b)}
\end{overpic}    
    \caption{(a) Equal-time spin structure factor $S(\bm{q})$ from DMRG [see Eq.~(\ref{eq:ssf_equal_time})] in the $[hhl]$-plane (upper panel) and in the $[hl0]$-plane (lower panel) for different values of $J_2$ for the 32-site cluster using 12000 SU(2) states. Due to the high values of the peaks at ${\bm q}=(4\pi,0,0)$ in the bottom right subfigure (\emph{cf.} Fig. \ref{fig:weight}), a cutoff was applied according to the colorbar to enhance the visibility of the other subfigures.  (b) Patterns of nearest neighbor spin correlations in the three degenerate ground states of the non-magnetic phase at $J_2\lesssim0.02$ and of the ${\bm k}=0$ magnetic phase at $J_2\gtrsim0.02$ as found by DMRG. In the non-magnetic phase blue bonds illustrate strong antiferromagnetic (AFM) correlations ($\left\langle \bm{S}_i\bm{S}_j \right\rangle_{\mathrm{strong}}\sim -1.42$ for $J_2=0.0125$ in DMRG) while red bonds denote weak correlations ($\left\langle \bm{S}_i\bm{S}_j \right\rangle_{\mathrm{weak}}\sim -0.058$ for $J_2=0.0125$ in DMRG). On the other hand, in the ${\bm k}=0$ magnetic phase blue bonds illustrate ferromagnetic (FM) correlations ($\left\langle \bm{S}_i\bm{S}_j \right\rangle_{\mathrm{FM}}\sim 0.75$ for $J_2=0.05$ in DMRG) while red bonds denote antiferromagnetic correlations ($\left\langle \bm{S}_i\bm{S}_j \right\rangle_{\mathrm{AFM}}\sim -1.13$ for $J_2=0.05$ in DMRG).}
        \label{fig:structure_factors}
\end{figure}

First, we investigate the system's ground-state properties using the DMRG technique. The extrapolated energies and the spin gap are shown in Fig. \ref{fig:energies}.
The ground-state energies exhibit a remarkably similar behavior to the $S=1/2$ case \cite{astrakhantsev_broken-symmetry_2021}: A shallow maximum emerges as $J_2$ is increased for both cluster sizes. This increase of the ground-state energy upon switching on $J_2$ indicates that small second neighbor couplings first {\it increase} the system's frustration. A further increase of $J_2$ leads to a decreasing ground-state energy which signals a reduction of frustration. Similarly, near the maximum a significant drop in the spin gap is observed. Based on this observation, we may suspect that the ground-state wave function undergoes a substantial change near $J_2\sim 0.02$. This is further confirmed by the equal-time spin structure factor $S(\bm{q})$, [see Eq.~(\ref{eq:ssf_equal_time})] shown in Fig.~\ref{fig:structure_factors}(a).
One sees that for small $J_2\lesssim0.02$ the equal-time spin structure factor $S(\bm{q})$ in the $[hl0]$ plane depends only weakly on $q_y$ [left panel of Fig.~\ref{fig:structure_factors}(a)], which was also observed for $J_2=0$ in a previous work~\cite{hagymasi2022prb}. The small dependence on $p_y$ suggests that the system consists of weakly correlated $x$-$z$ planes. An analysis of nearest neighbor correlations reveals that as for $J_2=0$ in Ref.~\cite{hagymasi2022prb} the system forms lines of strongly antiferromagnetically correlated spins along the edges of the tetrahedra in the $x$-$z$ plane. On the other hand, the nearest neighbor bonds connecting these lines are only weakly correlated resulting in effectively decoupled $x$-$z$ planes. These strongly and weakly correlated nearest neighbor spins are illustrated by blue and red bonds in Fig.~\ref{fig:structure_factors}(b), respectively, where the middle figure corresponds to the state with decoupled $x$-$z$ planes. The existence of a plane of strong correlations indicates a phase with broken $C_3$ rotation symmetry around the $[111]$ axis. Thus, the ground state in this small-$J_2$ regime is three-fold degenerate corresponding to the selection of either the $x$-$y$, $x$-$z$ or $y$-$z$ plane. The DMRG calculation converges to one of these states, depending on the random initial state. The other two partner states can be found by consecutive optimizations with the additional constraint that the state being optimized should be orthogonal to the previously optimized states~\cite{hagymasi2022prb}.

Increasing $J_2$ we first observe an increasing dependence on $q_y$ [middle panel of Fig.~\ref{fig:structure_factors}(a)] indicating growing correlations between $x$-$z$ planes. For $J_2\gtrsim0.02$, however, the pattern changes drastically with weight accumulating at the wave vector ${\bm q}=(4\pi,0,0)$ [right panel of Fig.~\ref{fig:structure_factors}(a)]. We identify this as the onset of $\bm{k}=0$ magnetic order. At this point several remarks are in order. As we consider a finite system, spontaneous magnetization can not occur, which is also consistent with the finite spin gap. Moreover, a real phase transition should only emerge in the thermodynamic limit, however, this drastic relocation of the weight is a clear precursor of a phase transition at $J_2\sim 0.02$.

Interestingly, the three-fold degeneracy of the non-magnetic ground state observed for $J_2\lesssim 0.02$ remains in the $\bm{k}=0$ regime at $J_2\gtrsim 0.02$. Furthermore, the correlations in these three states have a similar line-like pattern as in the non-magnetic phase but with different signs and strengths: The three states exhibit non-intersecting lines of ferromagnetically correlated spins in either the $x$-$y$, $x$-$z$ or $y$-$z$ plane while nearest neighbor correlations between such lines are antiferromagnetic.  Figure~\ref{fig:structure_factors}(b) also visualizes these states, where the blue (red) lines now correspond to ferromagnetic (antiferromagnetic) bonds. The three states produce peaks in the spin structure factor at the wave vectors ${\bm q}=(0,0,4\pi)$, $(0,4\pi,0)$ and $(4\pi,0,0)$, respectively. Again, the selection depends on the initial state in DMRG.  

\begin{figure}[!t]
    \centering
    \includegraphics[width=\columnwidth]{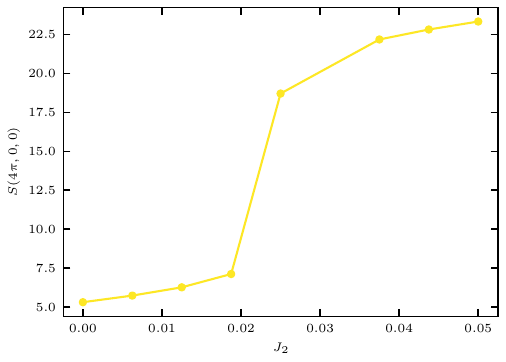}
    \caption{Spin structure factor $S({\bm q}=(4\pi,0,0))$ from DMRG at the ordering wave vector of the ${\bm k}=0$ phase as a function of $J_2$ for the 32-site cluster, using 12000 SU(2) states.}
        \label{fig:weight}
\end{figure}

To put the transition at $J_2\sim0.02$ on a more quantitative footing, the weight $S({\bm q}=(4\pi,0,0))$ at the ordering wave vector is plotted as a function of $J_2$ in Fig.~\ref{fig:weight}.
One can clearly observe a significant increase in signal near $J_2\sim 0.02$, which marks the phase boundary in the finite system.

A natural question is how representative our results are for the thermodynamic limit. Unfortunately, due to the enhanced growth of the Hilbert space compared to the $S=1/2$ case, it is not possible to consider larger clusters reliably, thus a finite-size scaling of the quantities discussed above is not possible. To address this issue and to corroborate our findings from DMRG, we use the complementary PMFRG approach, which is less constrained by finite size effects \cite{mueller2023pseudofermion,Niggemann2021a}.
\begin{figure}
    \centering
   \includegraphics[width=1\linewidth]{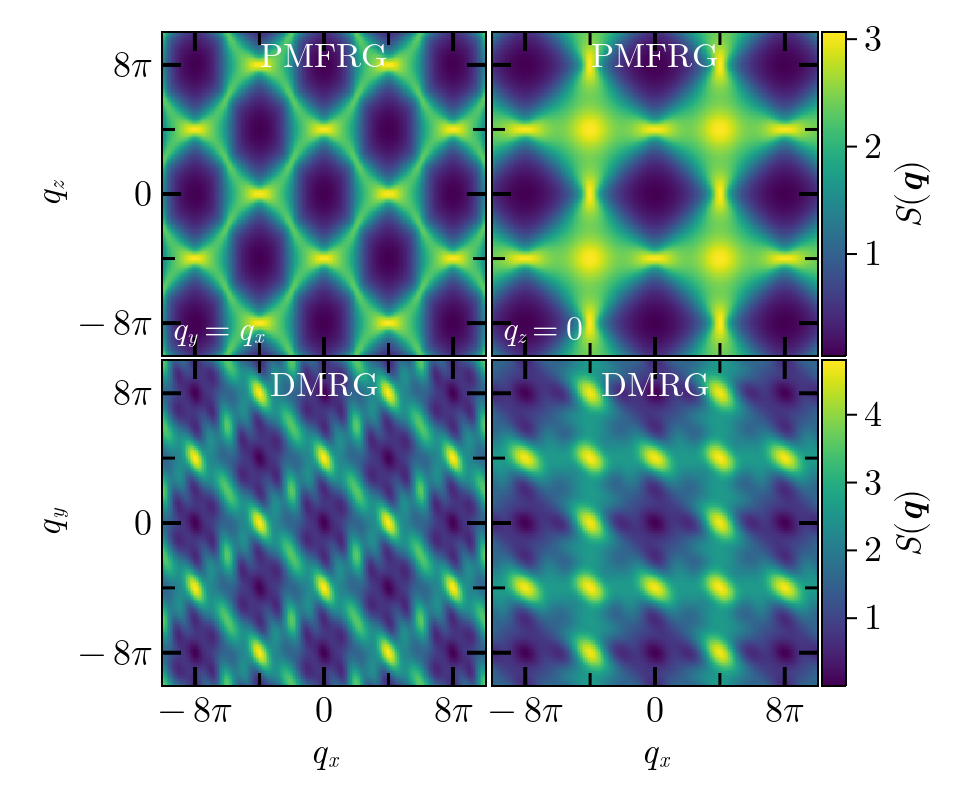}
    \caption{Spin structure factor $S(\bm q)$ for $J_2= 0.01875$ from DMRG at $T=0$ with symmetrized wavefunction (bottom panel) and from PMFRG at $T=0.25$ (top panel) for $L=10$. The left (right) plots show $S(\bm q)$ in the $[hhl]$ ($[hl0]$) plane.}
    \label{fig:Sq_comparison}
\end{figure}

\Cref{fig:Sq_comparison} shows the equal-time spin structure factor $S(\bm q)$ in the non-magnetic phase at $J_2=0.01875$ in comparison between PMFRG and DMRG. Since, by construction, all symmetries of the Hamiltonian remain intact within PMFRG, the DMRG data in Fig.~\ref{fig:Sq_comparison} is symmetrized to enable a direct comparison between both approaches. Also note that the DMRG result corresponds to $T=0$ while the PMFRG data is taken at the lowest simulated temperature $T=0.25$. Both methods agree well on their prediction of dominant wave vectors, however, in DMRG the peaks in $S(\bm q)$ (which indicate the proximity to the $\bm k=0$ ordered phase) are somewhat more pronounced. This is likely an effect of the different temperatures involved in the comparison, where the finite temperature $T=0.25$ used in PMFRG smears the signal. 
On the other hand, PMFRG can efficiently describe fully translationally invariant systems, restricting only the length of spin correlations (here, $L=10$). If $L$ is chosen to be larger than the physical correlation length, this method provides a virtually infinite momentum resolution, which smoothes the more `grainy' appearance of the spin structure factor from DMRG. As a consequence, remnants of the characteristic pinch-point pattern known from the $J_1$-only model become visible at $\bm q = (4\pi,0,0)$ within PMFRG. However, due to the finite $J_2$ coupling a peak starts to grow out of the pinch-point center~\cite{iqbal_quantum_2019}.

\begin{figure}
    \centering
    \includegraphics[width = \linewidth]{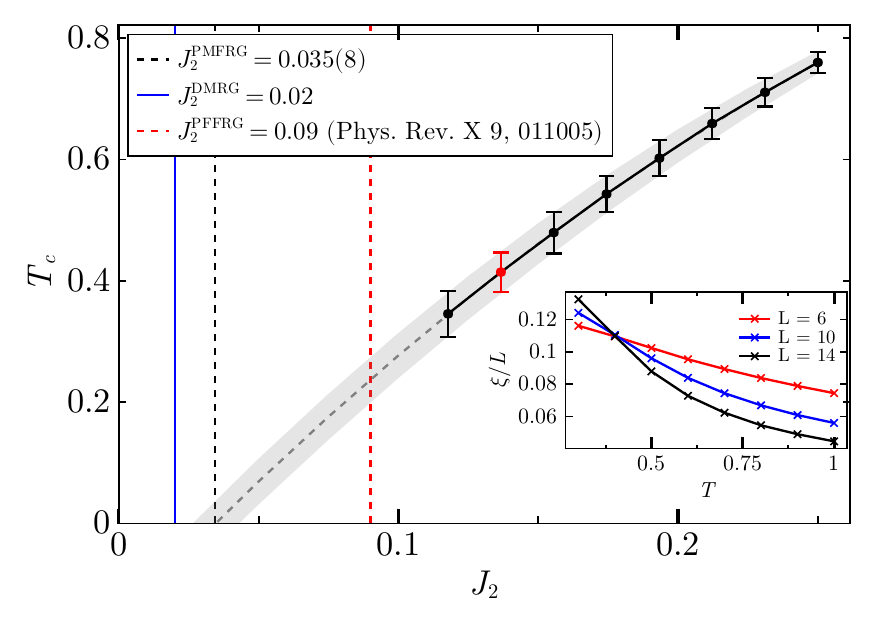}
    \caption{$J_2$ dependence of the critical temperature $T_c$, obtained via a scaling collapse of the correlation length $\xi \sim L$ calculated with one-loop PMFRG. As an example, the inset shows the scaling collapse for the red data point in the main panel for maximal correlation distances $L=6,10,14$. Error bars are estimated from the maximum deviation between crossing points of different curves. The gray dashed line is an extrapolation of the phase transition to $T=0$ using a parabolic fit, where the vertical black dashed line highlights the $T=0$ critical $J_2$ coupling obtained this way. The vertical blue line indicates the critical $J_2$ coupling from DMRG and the vertical red dashed line is the result from Ref.~\cite{iqbal_quantum_2019}.}
    \label{fig:PMFRG_tc}
\end{figure}

Upon increasing $J_2$ we observe growing intensities of the spin structure factor at the dominant wave vector $\bm q = (4\pi,0,0)$ within PMFRG (not shown). In order to rigorously determine the transition to the $\bm k=0$ ordered phase in PMFRG we study the model for three different values of the maximal correlation length $L = 6,10,14$ using the one-loop PMFRG scheme which was found to be more suitable for resolving magnetic order than higher loop-orders~\cite{niggemann2021quantitative}. We find critical temperatures of a second order phase transition via finite-size scaling of the correlation length $\xi$ [see Eq.~(\ref{eq:correlationRatio})] which, at the critical temperature $T_c$, scales as $\xi/L = \text{const.}$, i.e., linear in the numerically chosen maximum correlation distance $L$. The critical temperatures $T_c$ determined this way are shown in Fig.~\ref{fig:PMFRG_tc} as a function of $J_2$. Although temperatures below $T \lesssim 0.3$ cannot be resolved accurately due to the effects of neglecting higher order vertices in the flow equations, $T_c$ shows a clear downward trend when lowering $J_2$. The phase boundary $T_c(J_2)$ can be accurately fitted by a parabolic curve which we extrapolate to obtain an estimate for the phase boundary at $T=0$, given by $J_2=0.035(8)$. The error bars for the data points in \cref{fig:PMFRG_tc} are estimated from the maximal difference between pairwise crossing points of curves for different $L$ (see inset). The error of the $T=0$ phase boundary is estimated through extrapolations of the errorbars as displayed in the figure.

While the value $J_2=0.035(8)$ from PMFRG is somewhat larger than the critical coupling $J_2\sim0.02$ estimated with DMRG, taking into account the uncertainties of the extrapolation to $T=0$ in PMFRG and the fact that the DMRG result is obtained for a small spin cluster, both critical couplings can still be regarded as consistent. Particularly, the small critical $J_2$ found in both approaches indicates a high fragility of the non-magnetic phase of the nearest neighbor model to second neighbor couplings. 

\subsection{Discussion of previous PFFRG results}\label{sec:comparison}
The position of the zero-temperature phase transition at $J_2=0.035(8)$ identified by PMFRG in the previous subsection differs considerably from the critical coupling $J_2 = 0.09(2)$ found in Ref.~\cite{iqbal_quantum_2019} obtained by the similar PFFRG approach. This raises the question why these closely related approaches disagree so strongly in that result. The following discussion provides an explanation, points out precautions when using the PFFRG and describes how these methods should best be applied.

To start, it is worth explaining several properties of the PFFRG and how this method detects quantum phase transitions between magnetically ordered and disordered phases. The PFFRG expresses spin operators in terms of {\it complex fermionic} auxiliary particles, so-called Abrikosov fermions, which introduce unphysical states. While at $T=0$, the impact of unphysical states is usually found to be mild, at finite temperatures their contributions grow such that a meaningful application of the PFFRG is restricted to $T=0$ (unless a projection via the Popov-Fedotov method is applied~\cite{schneider2022_taming}). Thus, a finite-temperature phase transition can never be directly observed in PFFRG but instead reveals itself at finite cutoff $\Lambda>0$ at $T=0$. However, while $\Lambda$ shares some properties with the temperature $T$, its artificial nature complicates the physical interpretation of results at finite $\Lambda$. For example, it is currently unclear whether critical scaling in the system size $L$ is generally expected at finite $\Lambda$. Hence, the usual approach to nevertheless identify the onset of magnetic order for given coupling parameters at finite $\Lambda$ and $T=0$ within PFFRG is to search for instability features of the susceptibility as a function of $\Lambda$, such as kinks or peaks. This identification of long-range order, however, gets increasingly difficult if one approaches a quantum critical point from the ordered side, since kinks get less pronounced, shift to lower $\Lambda$ and continuously disappear. Furthermore, at low $\Lambda$ the PFFRG becomes increasingly inaccurate (for the same intrinsic methodological reason why PMFRG becomes uncontrolled at low $T$) and the existence or absence of weak instability features may sensitively depend on details of the implementation such as the chosen frequency mesh. As an example, Fig.~13 of Ref.~\cite{iqbal_quantum_2019} illustrates these difficulties in precisely locating zero-temperature phase boundaries in PFFRG, where a kink in the susceptibility flow of the nearest neighbor pyrochlore Heisenberg model continuously disappears when the spin magnitude $S$ is decreased. Usually, in PFFRG the extent of a magnetically ordered zero-temperature phase is determined by the coupling parameter regime in which kinks in the $\Lambda$ flow of the susceptibility are explicitly visible (although sometimes only faintly visible). The phase boundaries obtained this way (such as the critical second neighbor coupling $J_2 = 0.09(2)$ of the $S=1$ $J_1$-$J_2$ pyrochlore Heisenberg model obtained in Ref.~\cite{iqbal_quantum_2019}), might, however, underestimate the extent of ordered phases because there could be a parameter window where magnetic order exists, but the instability features are too faint to be visible in the susceptibility or the methodological limitations at small $\Lambda$, coarse frequency meshes and finite system sizes prevent their observation.

Let us compare this situation with the more quantitative method of locating a $T=0$ phase transition in PMFRG where a finite size scaling and an extrapolation to $T=0$ is employed. The phase diagram in Fig.~\ref{fig:PMFRG_tc} shows that a phase transition into a ${\bm k}=0$ magnetic regime is only explicitly observed via critical scaling for $J_2\gtrsim 0.12$, while for smaller $J_2$ the methodological limitations at low $T$ or the actual absence of order prevent a direct detection of a phase transition. If the extent of the ${\bm k}=0$ ordered phase is only determined from the $J_2$ regime where a transition is directly visible (via a susceptibility kink in PFFRG or via critical scaling in PMFRG) the two values $J_2 = 0.09(2)$ in PFFRG and $J_2= 0.12$ in PMFRG are actually not so different. This indicates that the competition between magnetic order phenomena and magnetic disorder fluctuations might be described quite similarly in both approaches. A crucial difference, however, is that the more quantitative approach of detecting phase transitions in PMFRG allows an extrapolation to $T=0$, and thus provides access to actual quantum phase transitions. Note that an analogous procedure in PFFRG, i.e., an extrapolation of kink-like instability features to $\Lambda=0$ would be considerably harder, since the $\Lambda$ positions of weak kinks often depend sensitively on details of the implementation (precise choice of the frequency grid).

To summarize this discussion, the much larger zero-temperature critical $J_2$ coupling from PFFRG compared to PMFRG does not necessarily indicate that PFFRG intrinsically underestimates magnetic order. Rather, the usual approach of identifying a quantum phase transition from PFFRG susceptibility data is inaccurate and can make magnetically ordered phases appear smaller than they actually are. Therefore, zero-temperature phase boundaries from PFFRG are only rough estimates and the obtained sizes of magnetically ordered regimes can be understood as lower bounds for the actual extents of these phases. The results in this paper also show how PMFRG solves this problem via extrapolations of phase boundaries to $T=0$, giving rise to more accurate positions of quantum phase transitions.

\section{Conclusions}\label{sec:conclusion}
We investigated the $S=1$ antiferromagnetic Heisenberg model on the pyrochlore lattice with first ($J_1$) and second neighbor ($J_2$) interactions. Our studies make use of matrix-product-state based techniques and PMFRG, two very different approaches with complementary strengths and capabilities. While DMRG is restricted to finite spin clusters, it has direct access to actual wave functions and is applicable at $T=0$. On the other hand, the PMFRG can approach the thermodynamic limit, however, it is restricted in the quantities it can calculate (mostly spin correlations) and becomes less accurate at low temperatures. Combining the different strengths of both approaches we develop a consistent phase diagram of the $S=1$ $J_1$-$J_2$ pyrochlore Heisenberg model. For small $J_2\lesssim0.02$, the system shows non-magnetic behavior. In this regime, DMRG finds that the three-fold degeneracy of the ground state that was previously identified at $J_2=0$ persists at finite $J_2$. Furthermore, the system is found to exhibit weakly coupled spin chains located in one of the three cubic faces, breaking the three-fold rotation symmetry around the $[111]$ axis. Investigating this phase with PMFRG, the large accessible system sizes and enhanced momentum resolution allow us to observe remnants of pinch-points in the spin structure factor, which, however, due to the finite $J_2$ coupling, are superimposed by a peak indicating the proximity to an ordered phase. Around $J_2\sim0.02$ DMRG finds a drastic change of the spin correlations indicated by the emergence of a sharp peak in the spin structure factor at ${\bm q}=(4\pi,0,0)$ suggesting the onset of ${\bm k}=0$ magnetic order. PMFRG confirms this phase transition in a similar regime at $J_2=0.035(8)$ by extrapolating a finite temperature transition to $T=0$. Interestingly, our DMRG results in the magnetically ordered phase also indicate a three-fold ground state degeneracy and line-like patterns of spin correlations in one of the cubic faces, similar to the non-magnetic phase.

Besides the physical insights into the $S=1$ antiferromagnetic Heisenberg model on the pyrochlore lattice which our paper provides, we also discuss several important methodological aspects of the applied methods. Particularly, our studies demonstrate that the (unavoidable) unphysical spin states of a $S=1$ pseudo Majorana model can be effectively eliminated in PMFRG via level repulsion terms. The accuracy of this approach is confirmed by the excellent agreement with finite temperature heat capacity and susceptibility data at $J_2=0$ obtained by the density-matrix purification approach. Finally, we explain discrepancies with earlier PFFRG results for the same model and argue that our current findings can be considered more accurate.

\begin{acknowledgments}
We would like to thank the participants of the spin-FRG workshop 2023 in Berlin for stimulating discussion.
I.\,H.~was supported in part by the Hungarian National Research,   
Development   and   Innovation Office (NKFIH) through Grants No.~K120569 and No.~K134983 and by the Quantum Information National Laboratory of Hungary.
N.~N.~and J.~R.~acknowledge support from the Deutsche
Forschungsgemeinschaft (DFG, German Research Foundation), within Project-ID 277101999 CRC 183 (Project
A04).
We acknowledge the use of the JUWELS cluster at the Forschungszentrum J\"ulich and the Noctua2 cluster at the Paderborn Center for Parallel Computing (PC$^2$). Some of the data presented here was produced using the \textsc{SyTen} toolkit~\cite{hubig:_syten_toolk,hubig17:_symmet_protec_tensor_networ}.
\end{acknowledgments}

\renewcommand{\thesection}{Appendix}
\renewcommand{\theequation}{A\arabic{equation}}

\section{Generalization of PMFRG to arbitrary spin magnitudes $S$}
\label{app:Appendix}

\subsection{Spin representation}
\label{app:spinRep}
Although the implementation of higher spin magnitudes $S$ was previously shown in the context of PFFRG~\cite{Baez17}, the present case requires further considerations which we now discuss in detail.
The SO(3) Majorana representation employed in PMFRG is applicable only for spin-$1/2$ operators. Ideally, the solution for $S>1/2$ would be to find a representation of spin-$S$ operators in terms of Majorana fermions, which does not introduce any unphysical states. However, such a representation exists only for $S=1/2$ and $S=3/2$, whereas for all other spin magnitudes unphysical sectors cannot be avoided~\cite{schaden2023bilinear}. 
Thus, in the present $S=1$ case we follow the approach of Ref.~\cite{Baez17} that introduces various copies (replicas) of spin-$1/2$ degrees of freedom on each site but that also inevitably involves unphysical states that need to be dealt with. 

Specifically, our approach of implementing an effective spin quantum number $S_\text{eff}$ amounts to introducing $2S_\text{eff}$ spin-$1/2$ operators $\bm S_{i_\mu}$ on each site $i$ such that 
\begin{equation}
    \bm S_{i,\text{eff}} = \sum_{\mu=1}^{2S_\text{eff}} \bm S_{i_\mu}, \label{eq:S_eff}
\end{equation}
where $\mu\in\{1,2,\ldots,2S_\text{eff}\}$ is the additional replica index. Here and in the following, we use the convention that $\bm S_{i_\mu}$ refers to a spin-$1/2$ operator, corresponding to the $\mu$-th replica spin located on site $i$ whereas operators that implement higher spins $S_\text{eff}$ are denoted $\bm S_{i,\text{eff}}$.

Thus, the effective model treated in PMFRG model results from replacing spin operators by ${\bm S}_{i,\text{eff}}$ according to Eq.~(\ref{eq:S_eff}) giving rise to a Hamiltonian in terms of spin-$1/2$ operators,
\begin{align}\label{eq_SpinSHam}
	H &= \sum_{(i, j)}\sum_{\mu, \nu}J_{ij} \bm{S}_{i_\mu} \cdot \bm{S}_{j_\nu} + A \sum_i \left( \sum_\mu \bm{S}_{i_\mu} \right)^2\nonumber\\
	&\equiv \sum_{(i, j)}\sum_{\mu, \nu}J_{i_\mu;j_\nu} \bm{S}_{i_\mu} \cdot \bm{S}_{j_\nu},
\end{align}
where $(i,j)$ denotes pairs of sites $i\neq j$ (summed over only once). To tune the energy of unphysical states, we add a level repulsion term $\sim A$ on the right hand side of the first line of Eq.~(\ref{eq_SpinSHam}): For $A<0$, the Hilbert space sector where the addition of angular momenta realizes the largest spin quantum number $S_\text{eff}$ is energetically favored over (unphysical) states with smaller spin quantum numbers $S<S_\text{eff}$. Furthermore, in the second line of Eq.~(\ref{eq_SpinSHam}) we combine the original couplings $J_{ij}$ and the level repulsion $A$ into a joint interaction constant $J_{i_\mu;j_\nu}$ which now depends on site indices $i,j$ and replica indices $\mu,\nu$.

\begin{figure}
    \centering
    \includegraphics[width = 0.4\linewidth]{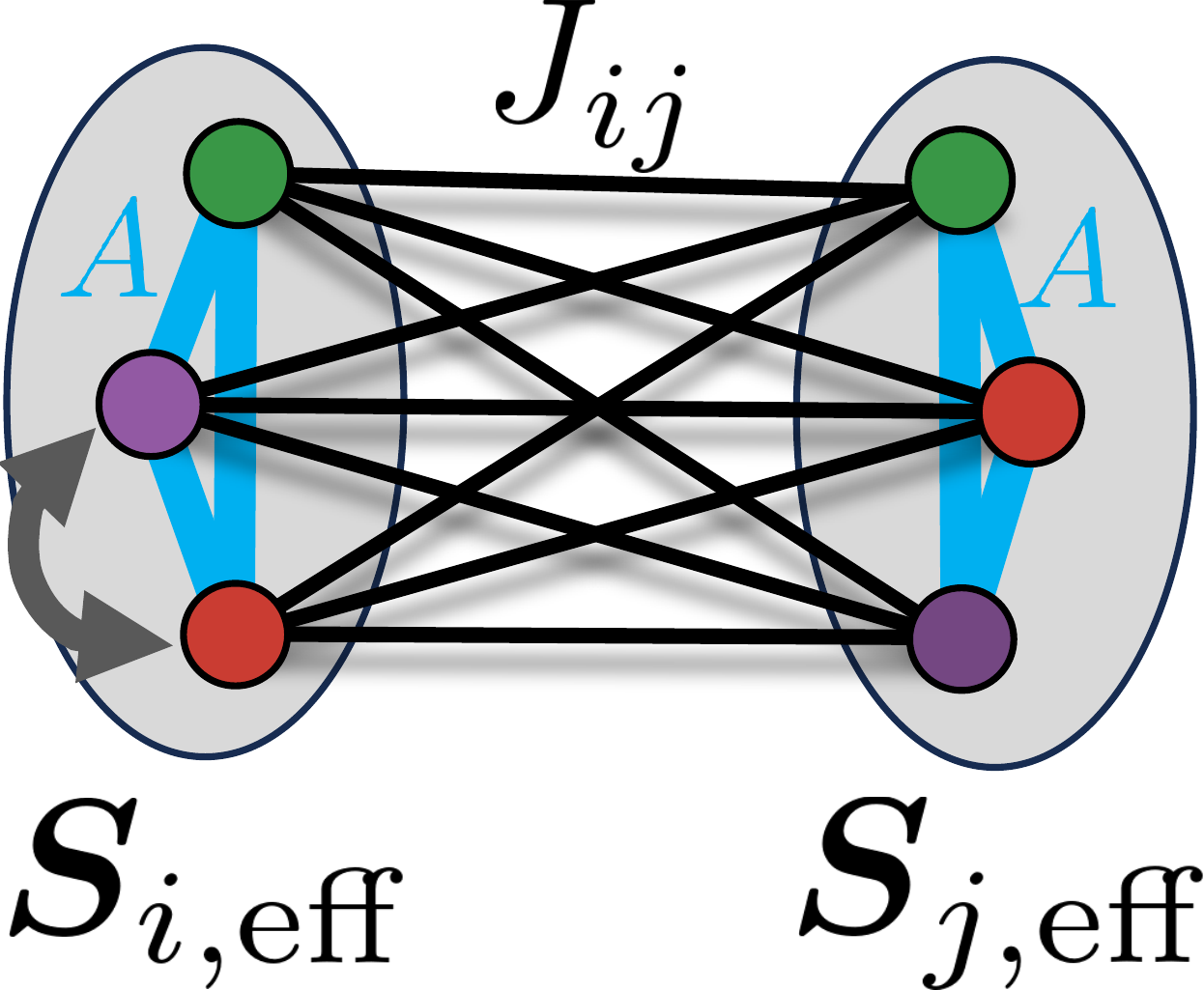}
    \caption{Illustration of two effective spin-$3/2$ degrees of freedom, each composed of three spin-$1/2$ replicas, interacting with each other. The spin replicas $S_{i_1}, S_{i_2}, S_{i_3}$ are fully equivalent and colored differently only for visual clarity. The grey arrow indicates an exemplary permutation of replicas which leaves the model invariant.}
    \label{fig:spinSPMFRG}
\end{figure}

The replica construction is depicted in \cref{fig:spinSPMFRG} for the case of two interacting $S_\text{eff}=3/2$ spins. Colored circles correspond to $S=1/2$ replicas which are coupled to each other locally with the ferromagnetic coupling $A$ and non-locally via the exchange interaction $J_{ij} $.
Strictly speaking, this mapping is only exact in the limit $A \rightarrow -\infty$, which, however, cannot be treated within PMFRG. This is due to the fact that the approximation of neglecting higher vertices in PMFRG breaks down if the effective interaction becomes too large compared to the temperature. In practice, it is sufficient to choose $A$ to be proportional to the temperature, $A = -\gamma T$, such that unphysical states are always gapped out, while $A$ is still small enough to avoid methodological challenges. Below, we will determine the best parameter $\gamma$ in the present $S=1$ case.

\subsection{Flow equations and correlation functions}
\label{app:FlowEq}
In PMFRG, we thus simulate the model in Eq.~(\ref{eq_SpinSHam}), i.e., a system of $2S_\textrm{eff} N$ spin-1/2 degrees of freedom $\bm{S}_{i_\mu}$ that are labelled by a site index $i$ and a replica index $\mu$. At first glance treating this model may look like a considerable increase in numeric complexity since the number of bonds with a coupling $J_{ij}$ increases by a factor $(2S_\textrm{eff})^2$ relative to a genuine spin-$1/2$ model in which the replica index takes only a single value $\mu=1$ (i.e., no different replicas exist). 
However, below we will see that the introduction of replicas has virtually no impact on numerical performance due to a large number of symmetries in the replica index.

The central object to be calculated in PMFRG is the vertex function $\Gamma_{f; i_\mu;j_\nu}(s,t,u)$ corresponding to the effective interaction between sites/replicas $i_\mu$ and $j_\nu$, where $f$ corresponds to the flavor combination, e.g. $xyxy$~\cite{niggemann2021majorana} and $s,t,u$ are Matsubara frequencies. At the initial cutoff scale $\Lambda\rightarrow\infty$ the vertex function is determined by the bare couplings $J_{i_\mu;j_\nu}$. Since an efficient PMFRG implementation only considers {\it symmetry-inequivalent} vertex functions $\Gamma_{f; i_\mu;j_\nu}(s,t,u)$, the first step is to determine all independent site/replica arguments $(i_\mu,j_\nu)$ that have to be taken into account.
As illustrated in~\cref{fig:spinSPMFRG}, replicas $\mu$, positioned on the same site $i$, are equivalent, implying an invariance of the system under local permutations of replicas $i_\mu \rightarrow i_{\mu'}$. Thus, the site/replica arguments of vertex functions $(i_\mu,j_\nu)$ can be divided into three equivalence classes of bonds denoted $(i_1,j_1)$, $(i_1,i_1)$, $(i_1,i_2)$ where each of the three bonds corresponds to a representative element in each class. The three classes are defined by
\begin{equation}
    (i_\mu,j_\nu) = \begin{cases}
         (i_1,j_1), \quad i \neq j\\
         (i_1,i_1), \quad i=j \textrm{ and } \mu = \nu \\
         (i_1,i_2),\quad i=j \textrm{ and } \mu \neq \nu
    \end{cases}.\label{eq:NewBonds}
\end{equation}
The first $(i_1,j_1)$ and second bond ($i_1,i_1$) are the inter-site and onsite bonds already known from a genuine spin-$1/2$ model. On the other hand, the last bond $(i_1,i_2)$ between two different replicas on the same physical site $i$ does not exist in the genuine spin-$1/2$ case.
Notably, apart from additional factors in the flow equation as discussed below, the introduction of replicas only amounts to the consideration of the additional symmetry-inequivalent bond $(i_1,i_2)$ regardless of the value of $S_\textrm{eff}$ which explains the negligible costs of our replica construction.

We now make use of the equivalence classes to determine the flow equations for arbitrary $S_\textrm{eff}$.
To this end, we consider an exemplary term on the right hand side of the flow equation for $d\Gamma_{f; ij}(s,t,u)/d\Lambda$ in Eq.~(51) of Ref.~\cite{niggemann2021majorana} which, before the introduction of replicas, has the form
\begin{align}
    X_{ij} =  \sum_k \GammaL_{ki} \GammaR_{kj} P_{kk}. \label{eq:Xij}
\end{align}
\Cref{eq:Xij} is written in a strongly condensed form: As frequency indices and Majorana flavors $x,y,z$ are irrelevant for our discussion, they are omitted in the bubble propagator $P_{ii} \equiv -S^\Lambda_i(\omega)G^\Lambda_i(\omega')$ where $S^\Lambda_i(\omega)$ is the single-scale propagator and $G^\Lambda_i(\omega')$ is the fully dressed single-particle propagator. Likewise, to suppress frequency and flavor indices, we use dummy labels for vertices $\GammaL_{ij}$ and $\GammaR_{ij}$ to refer to arbitrary left and right vertices as appearing in each term of the flow equations in Eq.~(51) of Ref.~\cite{niggemann2021majorana}, for example $\GammaL_{ij} = \Gamma^\Lambda_{a, ij}(s,\omega+\omega_1,\omega+\omega_2)$ in Eq.~(51a) of Ref.~\cite{niggemann2021majorana}.

Upon introducing the spin-$1/2$ replicas by splitting the sum over $k$ as $\sum_k \rightarrow \sum_k \sum^{2S_\textrm{eff}}_{\mu=1}$, we reorganize and group the terms in the summation over $\mu$ according to the equivalence classes in \cref{eq:NewBonds}.
We note that due to the aforementioned permutation symmetry in the replica index $\mu$, we may write the propagator as $G_{i_\mu} = G_{i_1} \equiv G_{i}$ and thus $P_{i_\mu,i_\mu} \equiv P_{i,i}$. Hence, Eq.~(\ref{eq:Xij}) in the class of bonds $(i_1,j_1)$ becomes
\begin{align}\label{eq:xi1j1}
    X_{i_1 j_1} &= \sum_{k\mu} \GammaL_{k_\mu,i_1} \GammaR_{k_\mu, j_1} P_{k_\mu,k_\mu} \nonumber\\
    &= 2S_\textrm{eff} \sum_{k \neq i,j} \GammaL_{k_1,i_1} \GammaR_{k_1, j_1} P_{k,k} \nonumber\\
    &+ \GammaL_{i_1,i_1} \GammaR_{i_1, j_1} P_{i,i} + \GammaL_{j_1,i_1} \GammaR_{j_1, j_1} P_{j,j}\nonumber\\
    &+ (2S_\textrm{eff}-1) \left( \GammaL_{i_1,i_2} \GammaR_{i_1, j_1} P_{i,i} + \GammaL_{j_1,i_1} \GammaR_{j_1, j_2} P_{j,j} \right).
\end{align}
The first term is equivalent to the non-local contributions in the genuine spin-$1/2$ case, rescaled by a factor $2S_\textrm{eff}$ accounting for all replicas. On the other hand, the newly introduced bond $(i_1,i_2)$ requires us to add the last line in Eq.~(\ref{eq:xi1j1}). It can be seen that in the special case $S_\textrm{eff} = 1/2$, when no replicas are introduced, this expression reduces back to \cref{eq:Xij}. 
In the same way we obtain for the bonds $(i_1,i_1)$ and $(i_1,i_2)$:
\begin{align}
    X_{i_1,i_1} &= 2S_\textrm{eff}  \sum_{k \neq i} \GammaL_{k_1,i_1} \GammaR_{k_1, i_1} P_{k,k}  \nonumber\\
    &+ \GammaL_{i_1,i_1} \GammaR_{i_1, i_1} P_{i,i} + (2S_\textrm{eff}-1) \left( \GammaL_{i_1,i_2} \GammaR_{i_1, i_2} P_{i,i} \right)
\end{align}
and
\begin{align}
    X_{i_1,i_2} &= 2S_\textrm{eff}  \sum_{k \neq i} \GammaL_{k_1,i_1} \GammaR_{k_1, i_1} P_{k,k}  \nonumber\\
    &+ \big(\GammaL_{i_1,i_1} \GammaR_{i_1, i_2} + \GammaL_{i_1,i_2} \GammaR_{i_1, i_1} \nonumber\\
    &+ (2S_\textrm{eff}-2) \GammaL_{i_1,i_2} \GammaR_{i_1, i_2} \big) P_{i,i}.
\end{align}
The site summation for the self energy similarly changes through the introduction of replicas and its terms, are the same as found for $X_{i_1,i_1}$, see \cite{niggemann2021majorana}.
It is worth emphasizing again that the only additional terms compared to a genuine spin-1/2 model are those containing vertex functions on $(i_1,i_2)$ bonds. Hence, for a PMFRG implementation of models with higher spins $S>1/2$ one can copy most terms from a spin-$1/2$ code while only the terms with $(i_1,i_2)$ bonds need to be added manually~\footnote{This must only be implemented once for an arbitrary lattice. An examplary code implementation can be found in our publicly available package \href{https://github.com/NilsNiggemann/SpinFRGLattices.jl/blob/dac0bc73ec6e28f82e9c445d8a5a78671ccb48b9/src/SpinSGeneralization.jl}{SpinFRGLattices.jl}, where a given lattice geometry may simply be modified to obtain the corresponding effective spin-$S$ model.}.

Ultimately, the object of interest are spin correlators $\chi_{ij} \equiv \langle \bm S_{i,\text{eff}}\cdot\bm S_{j,\text{eff}} \rangle $. Since the PMFRG only returns replica correlators $\langle \bm S_{i_\mu} \cdot\bm S_{j_\nu} \rangle$, we obtain $\chi_{ij}$ via replacing $\bm S_{i,\textrm{eff}}$ by its definition in \cref{eq:S_eff} and expanding the replica summation. Again, we use the equivalence classes in \cref{eq:NewBonds} to simplify the expression. For non-local correlators $i\neq j$, all terms in the sum are equal and thus we obtain a simple prefactor $(2S_\textrm{eff})^2$. For local correlators $i=j$ we distinguish between contributions from two different replicas  $\langle \bm S_{i_1} \cdot\bm S_{i_2} \rangle$ and contributions from identical replicas $\langle \bm S_{i_1} \cdot\bm S_{i_2} \rangle$:
\begin{align}
   \chi_{i,j\neq i} &=
   (2S_\textrm{eff})^2 \langle \bm S_{i_1} \cdot\bm S_{j_1} \rangle\\
  \chi_{i,i}  &=2S_\textrm{eff} \langle \bm S_{i_1} \cdot\bm S_{i_1} \rangle + 2S_\textrm{eff}(2S_\textrm{eff}-1) \langle \bm S_{i_1} \cdot\bm S_{i_2} \rangle
\end{align}
As the heat capacity is determined from PMFRG via derivatives of the interacting free energy, we also adjust this quantity to be
\begin{equation}
    f_\textrm{int} \rightarrow 2S_\textrm{eff} f_\textrm{int}.
\end{equation}
\subsection{Benchmarking}
\label{app:Benchmarking}
\begin{figure}
    \centering
    \includegraphics[width = 0.7\linewidth]{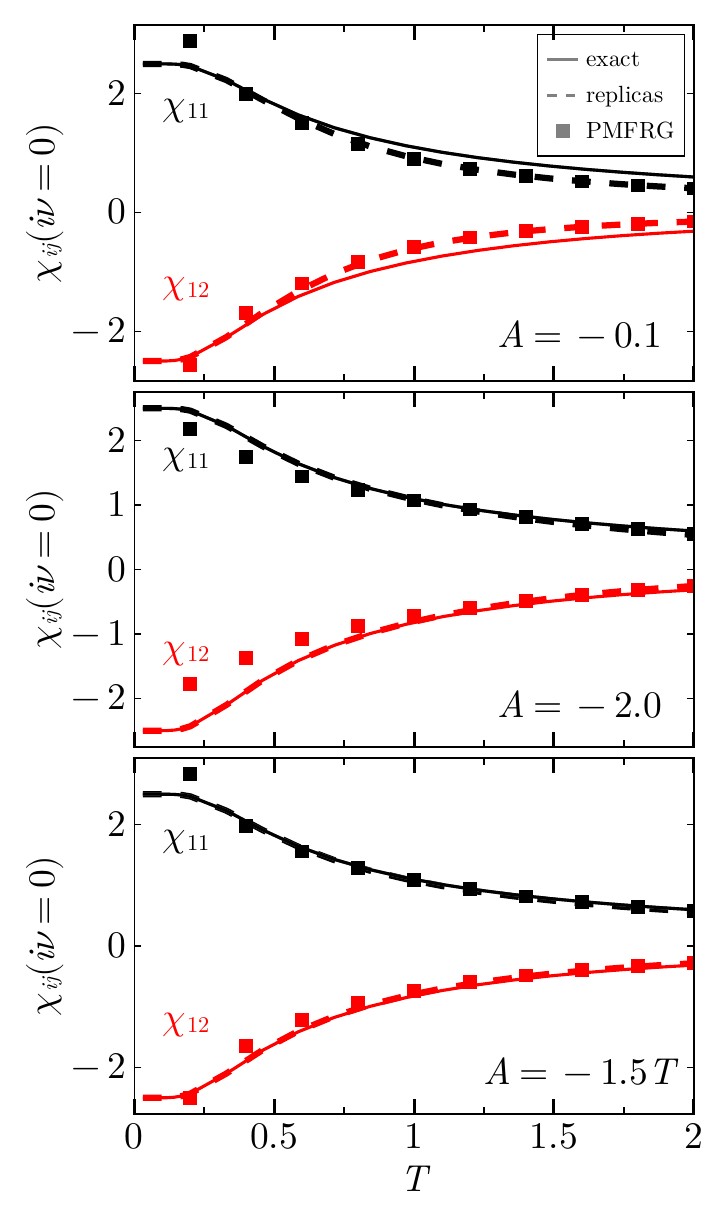}
    \caption{Spin correlations of the $S=3/2$ Heisenberg dimer for different choices of the level repulsion, $A=-0.1$, $A=-2$, $A=-1.5T$, from top to bottom. The solid black (red) line indicates the exact solution of the local correlator $\chi_{11}$ (non-local correlator $\chi_{12}$). The dashed lines correspond to the exact solution after the introduction of replica spins [see \cref{fig:spinSPMFRG}], i.e., in the presence of unphysical $S=1/2$ spin states. Squares indicate results from PMFRG.}
    \label{fig:S32Dimer_PMFRG}
\end{figure}

As a first check of our implementation, we consider the exactly solvable system $H= J\bm S_{1, \textrm{eff}}\cdot \bm S_{2, \textrm{eff}} $ of two interacting $S=3/2$ spins $\bm S_{1, \textrm{eff}}$ and $\bm S_{2, \textrm{eff}}$ with an antiferromagnetic Heisenberg coupling $J>1$. We note that despite its small Hilbert space, this model poses the same (or even larger) challenges to our diagrammatic approach than real many-body systems. In fact, PMFRG is most suited for systems of higher dimensions due to its inclusion of mean-field contributions \cite{muller2023pseudo}. Consequently, studying the two-site Heisenberg dimer is a suitable test for the worst-case performance of our method.
The Hamiltonian of this system treated via replicas $\mu=1,2,3$ is \begin{equation}
    H = J\sum^{3}_{\mu=1}\sum^{3}_{\nu=1} \bm S_{1_\mu} \cdot \bm S_{2_\nu} + A \sum^2_{i=1}\left(\sum^{3}_{\mu=1} \bm S_{i_\mu}\right)^2, \label{eq:DimerReplicas}
\end{equation}
where we set $J=1$ in the following. In \cref{fig:S32Dimer_PMFRG} we present results for the static local spin correlator $\chi_{11}(i\nu=0)$ and the static non-local spin correlator $\chi_{12}(i\nu=0)$ defined by
\begin{equation}
    \chi_{ij}(i\nu=0) = \int_0^\beta d\tau \langle \bm S^z_{i,\text{eff}}(\tau) \bm S^z_{j,\text{eff}}(0) \rangle,
\end{equation}
where $\tau$ is the imaginary time. Shown in \cref{fig:S32Dimer_PMFRG} are three different choices of the level repulsion $A$.
For small level repulsion $|A|=0.1 \ll J \equiv 1$, we observe good agreement between the exact results for the spin-3/2 dimer before and after the introduction of replicas (i.e., without and with unphysical $S=1/2$ states) at low temperatures. This is expected since the ground state of an unfrustrated spin system typically lies in the sector with maximal (effective) spin as the interaction energy $\sim \bm S_{1,\text{eff}} \cdot \bm S_{2,\text{eff}}$ is largest in this case.
It can be seen that PMFRG agrees remarkably well with the exact result of the replica system in \cref{eq:DimerReplicas} (dashed line). At higher temperatures, the effect of unphysical states becomes visible since the excitation gap to such states, determined by $A$, is too small to suppress their impact on physical observables. 
Conversely, for a larger level repulsion of $A=-2$, we observe good agreement between PMFRG and the exact result at higher temperatures as the unphysical states are further shifted to higher energies. However, at low temperatures, methodological difficulties arise because the dominant energy scale in the Hamiltonian (now given by $A$), also sets the temperature scale $T\sim |A|$ below which results can become inaccurate. The solution to this problem is to introduce a level repulsion $A=-\gamma T$ that scales linearly in temperature, as shown in the bottom panel of \cref{fig:S32Dimer_PMFRG}. There we find that the value $\gamma=1.5$ maximizes the agreement between PMFRG and the exact result both at high and low temperatures.
\begin{figure}
    \centering
    \includegraphics[width = \linewidth]{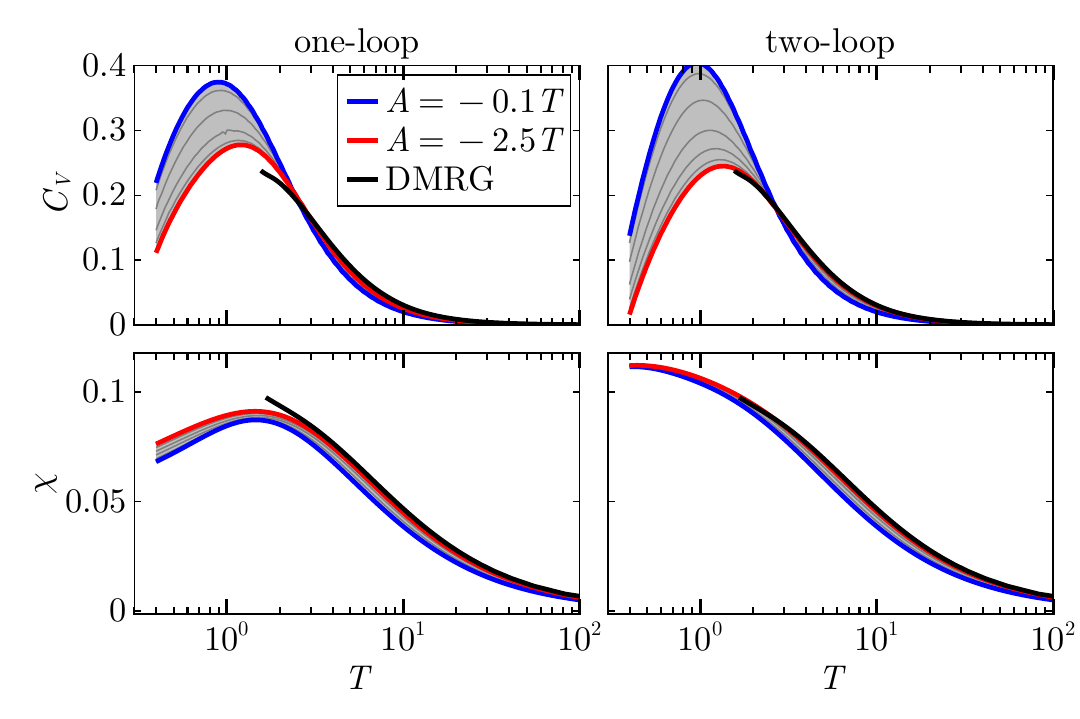}
    \caption{Specific heat (top panel) and uniform susceptibility (bottom panel) obtained from PMFRG in the standard one-loop truncation (left) and two-loop truncation (right) as a function of temperature for different choices of the level repulsion $A=-\gamma T$ ranging from $\gamma=0.1$ (blue lines) to $\gamma=2.5T$ (red lines). The values of $\gamma$ for the thin gray lines in between are $\gamma = 0.2,0.5,1,1.5,2$. The black lines denote the DMRG result.}
    \label{fig:DMRGComparison}
\end{figure}

Finally, we determine the optimal value $\gamma$ for the nearest neighbor pyrochlore model. In this case, we use DMRG data to benchmark our PMFRG results. Even though DMRG cannot access temperatures below $T\sim2$, the regime $T\gtrsim 2$ is still suitable for our benchmarks, since the effects of unphysical states are more pronounced at higher temperatures. In \cref{fig:DMRGComparison} we show the specific heat capacity $C_V$ and the uniform susceptibility $\chi$ a function of temperature for varying values of $\gamma$. At the smallest level repulsion $\gamma=0.1$, the effects of unphysical states are still clearly visible from the deviation between the PMFRG result (blue curve) and the DMRG result (black curve). As the level repulsion is increased, the PMFRG results first undergo significant changes, in particular the specific heat. However, for $\gamma\gtrsim 1$ the changes become smaller upon varying $\gamma$ and we observe convergence around $\gamma=2$. The agreement with DMRG is best at $\gamma=2.5$ (red curve) which is also used for our PMFRG results in the main text.
We note that for the conventional one-loop truncation, the specific heat and the susceptibility from PMFRG show a slightly enhanced deviation from the DMRG results at the lowest temperatures $T\sim2$ accessible within DMRG. As discussed in Ref.~\cite{niggemann2021quantitative}, this can be remedied by a two-loop truncation of the vertex flow equations, which significantly decreases these deviations.
In principle, the ideal choice of $\gamma$ can change upon increasing $J_2$. However the excitation gap to unphysical states at constant $\gamma$ and increasing $J_2$ is expected to increase as a result of the decreasing frustration. Hence, the constant value $\gamma=2.5$ should also properly eliminate the impact of unphysical states for $J_2>0$.

\bibliography{pyrochlore,pyrochlore_field,pyrochlore2trim}

\begin{thebibliography}{51}%
\makeatletter
\providecommand \@ifxundefined [1]{%
 \@ifx{#1\undefined}
}%
\providecommand \@ifnum [1]{%
 \ifnum #1\expandafter \@firstoftwo
 \else \expandafter \@secondoftwo
 \fi
}%
\providecommand \@ifx [1]{%
 \ifx #1\expandafter \@firstoftwo
 \else \expandafter \@secondoftwo
 \fi
}%
\providecommand \natexlab [1]{#1}%
\providecommand \enquote  [1]{``#1''}%
\providecommand \bibnamefont  [1]{#1}%
\providecommand \bibfnamefont [1]{#1}%
\providecommand \citenamefont [1]{#1}%
\providecommand \href@noop [0]{\@secondoftwo}%
\providecommand \href [0]{\begingroup \@sanitize@url \@href}%
\providecommand \@href[1]{\@@startlink{#1}\@@href}%
\providecommand \@@href[1]{\endgroup#1\@@endlink}%
\providecommand \@sanitize@url [0]{\catcode `\\12\catcode `\$12\catcode
  `\&12\catcode `\#12\catcode `\^12\catcode `\_12\catcode `\%12\relax}%
\providecommand \@@startlink[1]{}%
\providecommand \@@endlink[0]{}%
\providecommand \url  [0]{\begingroup\@sanitize@url \@url }%
\providecommand \@url [1]{\endgroup\@href {#1}{\urlprefix }}%
\providecommand \urlprefix  [0]{URL }%
\providecommand \Eprint [0]{\href }%
\providecommand \doibase [0]{https://doi.org/}%
\providecommand \selectlanguage [0]{\@gobble}%
\providecommand \bibinfo  [0]{\@secondoftwo}%
\providecommand \bibfield  [0]{\@secondoftwo}%
\providecommand \translation [1]{[#1]}%
\providecommand \BibitemOpen [0]{}%
\providecommand \bibitemStop [0]{}%
\providecommand \bibitemNoStop [0]{.\EOS\space}%
\providecommand \EOS [0]{\spacefactor3000\relax}%
\providecommand \BibitemShut  [1]{\csname bibitem#1\endcsname}%
\let\auto@bib@innerbib\@empty
\bibitem [{\citenamefont {Balents}(2010)}]{balents2010_spinliquid}%
  \BibitemOpen
  \bibfield  {author} {\bibinfo {author} {\bibfnamefont {L.}~\bibnamefont
  {Balents}},\ }\bibfield  {title} {\bibinfo {title} {Spin liquids in
  frustrated magnets},\ }\href {https://doi.org/10.1038/nature08917} {\bibfield
   {journal} {\bibinfo  {journal} {Nature}\ }\textbf {\bibinfo {volume}
  {464}},\ \bibinfo {pages} {199} (\bibinfo {year} {2010})}\BibitemShut
  {NoStop}%
\bibitem [{\citenamefont {Jiang}\ \emph {et~al.}(2008)\citenamefont {Jiang},
  \citenamefont {Weng},\ and\ \citenamefont {Sheng}}]{jiang_prl_2008}%
  \BibitemOpen
  \bibfield  {author} {\bibinfo {author} {\bibfnamefont {H.~C.}\ \bibnamefont
  {Jiang}}, \bibinfo {author} {\bibfnamefont {Z.~Y.}\ \bibnamefont {Weng}},\
  and\ \bibinfo {author} {\bibfnamefont {D.~N.}\ \bibnamefont {Sheng}},\
  }\bibfield  {title} {\bibinfo {title} {Density matrix renormalization group
  numerical study of the kagome antiferromagnet},\ }\href
  {https://doi.org/10.1103/PhysRevLett.101.117203} {\bibfield  {journal}
  {\bibinfo  {journal} {Phys. Rev. Lett.}\ }\textbf {\bibinfo {volume} {101}},\
  \bibinfo {pages} {117203} (\bibinfo {year} {2008})}\BibitemShut {NoStop}%
\bibitem [{\citenamefont {He}\ \emph {et~al.}(2017)\citenamefont {He},
  \citenamefont {Zaletel}, \citenamefont {Oshikawa},\ and\ \citenamefont
  {Pollmann}}]{pollmann_prx_2017}%
  \BibitemOpen
  \bibfield  {author} {\bibinfo {author} {\bibfnamefont {Y.-C.}\ \bibnamefont
  {He}}, \bibinfo {author} {\bibfnamefont {M.~P.}\ \bibnamefont {Zaletel}},
  \bibinfo {author} {\bibfnamefont {M.}~\bibnamefont {Oshikawa}},\ and\
  \bibinfo {author} {\bibfnamefont {F.}~\bibnamefont {Pollmann}},\ }\bibfield
  {title} {\bibinfo {title} {Signatures of dirac cones in a dmrg study of the
  kagome heisenberg model},\ }\href {https://doi.org/10.1103/PhysRevX.7.031020}
  {\bibfield  {journal} {\bibinfo  {journal} {Phys. Rev. X}\ }\textbf {\bibinfo
  {volume} {7}},\ \bibinfo {pages} {031020} (\bibinfo {year}
  {2017})}\BibitemShut {NoStop}%
\bibitem [{\citenamefont {Yan}\ \emph {et~al.}(2011)\citenamefont {Yan},
  \citenamefont {Huse},\ and\ \citenamefont {White}}]{yan_science_2011}%
  \BibitemOpen
  \bibfield  {author} {\bibinfo {author} {\bibfnamefont {S.}~\bibnamefont
  {Yan}}, \bibinfo {author} {\bibfnamefont {D.~A.}\ \bibnamefont {Huse}},\ and\
  \bibinfo {author} {\bibfnamefont {S.~R.}\ \bibnamefont {White}},\ }\bibfield
  {title} {\bibinfo {title} {Spin-liquid ground state of the s = 1/2 kagome
  heisenberg antiferromagnet},\ }\href
  {https://doi.org/10.1126/science.1201080} {\bibfield  {journal} {\bibinfo
  {journal} {Science}\ }\textbf {\bibinfo {volume} {332}},\ \bibinfo {pages}
  {1173} (\bibinfo {year} {2011})},\ \Eprint
  {https://arxiv.org/abs/https://science.sciencemag.org/content/332/6034/1173.full.pdf}
  {https://science.sciencemag.org/content/332/6034/1173.full.pdf} \BibitemShut
  {NoStop}%
\bibitem [{\citenamefont {Depenbrock}\ \emph {et~al.}(2012)\citenamefont
  {Depenbrock}, \citenamefont {McCulloch},\ and\ \citenamefont
  {Schollw\"ock}}]{schollwock_prl_2012}%
  \BibitemOpen
  \bibfield  {author} {\bibinfo {author} {\bibfnamefont {S.}~\bibnamefont
  {Depenbrock}}, \bibinfo {author} {\bibfnamefont {I.~P.}\ \bibnamefont
  {McCulloch}},\ and\ \bibinfo {author} {\bibfnamefont {U.}~\bibnamefont
  {Schollw\"ock}},\ }\bibfield  {title} {\bibinfo {title} {Nature of the
  spin-liquid ground state of the $s=1/2$ heisenberg model on the kagome
  lattice},\ }\href {https://doi.org/10.1103/PhysRevLett.109.067201} {\bibfield
   {journal} {\bibinfo  {journal} {Phys. Rev. Lett.}\ }\textbf {\bibinfo
  {volume} {109}},\ \bibinfo {pages} {067201} (\bibinfo {year}
  {2012})}\BibitemShut {NoStop}%
\bibitem [{\citenamefont {L\"auchli}\ \emph {et~al.}(2019)\citenamefont
  {L\"auchli}, \citenamefont {Sudan},\ and\ \citenamefont
  {Moessner}}]{lauchli_kagome_2019}%
  \BibitemOpen
  \bibfield  {author} {\bibinfo {author} {\bibfnamefont {A.~M.}\ \bibnamefont
  {L\"auchli}}, \bibinfo {author} {\bibfnamefont {J.}~\bibnamefont {Sudan}},\
  and\ \bibinfo {author} {\bibfnamefont {R.}~\bibnamefont {Moessner}},\
  }\bibfield  {title} {\bibinfo {title} {$s=\frac{1}{2}$ kagome heisenberg
  antiferromagnet revisited},\ }\href
  {https://doi.org/10.1103/PhysRevB.100.155142} {\bibfield  {journal} {\bibinfo
   {journal} {Phys. Rev. B}\ }\textbf {\bibinfo {volume} {100}},\ \bibinfo
  {pages} {155142} (\bibinfo {year} {2019})}\BibitemShut {NoStop}%
\bibitem [{\citenamefont {Kim}\ and\ \citenamefont {Han}(2008)}]{kim_prb_2008}%
  \BibitemOpen
  \bibfield  {author} {\bibinfo {author} {\bibfnamefont {J.~H.}\ \bibnamefont
  {Kim}}\ and\ \bibinfo {author} {\bibfnamefont {J.~H.}\ \bibnamefont {Han}},\
  }\bibfield  {title} {\bibinfo {title} {Chiral spin states in the pyrochlore
  heisenberg magnet: Fermionic mean-field theory and variational monte carlo
  calculations},\ }\href {https://doi.org/10.1103/PhysRevB.78.180410}
  {\bibfield  {journal} {\bibinfo  {journal} {Phys. Rev. B}\ }\textbf {\bibinfo
  {volume} {78}},\ \bibinfo {pages} {180410} (\bibinfo {year}
  {2008})}\BibitemShut {NoStop}%
\bibitem [{\citenamefont {Sobral}\ and\ \citenamefont
  {Lacroix}(1997)}]{sobral_1998}%
  \BibitemOpen
  \bibfield  {author} {\bibinfo {author} {\bibfnamefont {R.}~\bibnamefont
  {Sobral}}\ and\ \bibinfo {author} {\bibfnamefont {C.}~\bibnamefont
  {Lacroix}},\ }\bibfield  {title} {\bibinfo {title} {Order by disorder in the
  pyrochlore antiferromagnets},\ }\href
  {https://doi.org/https://doi.org/10.1016/S0038-1098(97)00212-3} {\bibfield
  {journal} {\bibinfo  {journal} {Solid State Communications}\ }\textbf
  {\bibinfo {volume} {103}},\ \bibinfo {pages} {407 } (\bibinfo {year}
  {1997})}\BibitemShut {NoStop}%
\bibitem [{\citenamefont {Canals}\ and\ \citenamefont
  {Lacroix}(2000)}]{canals_lacroix_prb_2000}%
  \BibitemOpen
  \bibfield  {author} {\bibinfo {author} {\bibfnamefont {B.}~\bibnamefont
  {Canals}}\ and\ \bibinfo {author} {\bibfnamefont {C.}~\bibnamefont
  {Lacroix}},\ }\bibfield  {title} {\bibinfo {title} {Quantum spin liquid: The
  heisenberg antiferromagnet on the three-dimensional pyrochlore lattice},\
  }\href {https://doi.org/10.1103/PhysRevB.61.1149} {\bibfield  {journal}
  {\bibinfo  {journal} {Phys. Rev. B}\ }\textbf {\bibinfo {volume} {61}},\
  \bibinfo {pages} {1149} (\bibinfo {year} {2000})}\BibitemShut {NoStop}%
\bibitem [{\citenamefont {Chandra}\ and\ \citenamefont
  {Sahoo}(2018)}]{candra_numerics_pyrochlore_correlations_2018}%
  \BibitemOpen
  \bibfield  {author} {\bibinfo {author} {\bibfnamefont {V.~R.}\ \bibnamefont
  {Chandra}}\ and\ \bibinfo {author} {\bibfnamefont {J.}~\bibnamefont
  {Sahoo}},\ }\bibfield  {title} {\bibinfo {title} {Spin-$\frac{1}{2}$
  heisenberg antiferromagnet on the pyrochlore lattice: An exact
  diagonalization study},\ }\href {https://doi.org/10.1103/PhysRevB.97.144407}
  {\bibfield  {journal} {\bibinfo  {journal} {Phys. Rev. B}\ }\textbf {\bibinfo
  {volume} {97}},\ \bibinfo {pages} {144407} (\bibinfo {year}
  {2018})}\BibitemShut {NoStop}%
\bibitem [{\citenamefont {Iqbal}\ \emph {et~al.}(2019)\citenamefont {Iqbal},
  \citenamefont {M{\"u}ller}, \citenamefont {Ghosh}, \citenamefont {Gingras},
  \citenamefont {Jeschke}, \citenamefont {Rachel}, \citenamefont {Reuther},\
  and\ \citenamefont {Thomale}}]{iqbal_quantum_2019}%
  \BibitemOpen
  \bibfield  {author} {\bibinfo {author} {\bibfnamefont {Y.}~\bibnamefont
  {Iqbal}}, \bibinfo {author} {\bibfnamefont {T.}~\bibnamefont {M{\"u}ller}},
  \bibinfo {author} {\bibfnamefont {P.}~\bibnamefont {Ghosh}}, \bibinfo
  {author} {\bibfnamefont {M.~J.~P.}\ \bibnamefont {Gingras}}, \bibinfo
  {author} {\bibfnamefont {H.~O.}\ \bibnamefont {Jeschke}}, \bibinfo {author}
  {\bibfnamefont {S.}~\bibnamefont {Rachel}}, \bibinfo {author} {\bibfnamefont
  {J.}~\bibnamefont {Reuther}},\ and\ \bibinfo {author} {\bibfnamefont
  {R.}~\bibnamefont {Thomale}},\ }\bibfield  {title} {\bibinfo {title} {Quantum
  and {Classical} {Phases} of the {Pyrochlore} {Heisenberg} {Model} with
  {Competing} {Interactions}},\ }\href
  {https://doi.org/10.1103/PhysRevX.9.011005} {\bibfield  {journal} {\bibinfo
  {journal} {Phys. Rev. X}\ }\textbf {\bibinfo {volume} {9}},\ \bibinfo {pages}
  {011005} (\bibinfo {year} {2019})}\BibitemShut {NoStop}%
\bibitem [{\citenamefont {M{\"u}ller}\ \emph {et~al.}(2019)\citenamefont
  {M{\"u}ller}, \citenamefont {Lohmann}, \citenamefont {Richter},\ and\
  \citenamefont {Derzhko}}]{muller_thermodynamics_2019}%
  \BibitemOpen
  \bibfield  {author} {\bibinfo {author} {\bibfnamefont {P.}~\bibnamefont
  {M{\"u}ller}}, \bibinfo {author} {\bibfnamefont {A.}~\bibnamefont {Lohmann}},
  \bibinfo {author} {\bibfnamefont {J.}~\bibnamefont {Richter}},\ and\ \bibinfo
  {author} {\bibfnamefont {O.}~\bibnamefont {Derzhko}},\ }\bibfield  {title}
  {\bibinfo {title} {Thermodynamics of the pyrochlore-lattice quantum
  {Heisenberg} antiferromagnet},\ }\href
  {https://doi.org/10.1103/PhysRevB.100.024424} {\bibfield  {journal} {\bibinfo
   {journal} {Phys. Rev. B}\ }\textbf {\bibinfo {volume} {100}},\ \bibinfo
  {pages} {024424} (\bibinfo {year} {2019})}\BibitemShut {NoStop}%
\bibitem [{\citenamefont {Tsunetsugu}(2017)}]{tsunetsugu_theory_2017}%
  \BibitemOpen
  \bibfield  {author} {\bibinfo {author} {\bibfnamefont {H.}~\bibnamefont
  {Tsunetsugu}},\ }\bibfield  {title} {\bibinfo {title} {Theory of
  antiferromagnetic {Heisenberg} spins on a breathing pyrochlore lattice},\
  }\bibfield  {journal} {\bibinfo  {journal} {Prog Theor Exp Phys}\ }\textbf
  {\bibinfo {volume} {2017}},\ \href {https://doi.org/10.1093/ptep/ptx023}
  {10.1093/ptep/ptx023} (\bibinfo {year} {2017})\BibitemShut {NoStop}%
\bibitem [{\citenamefont {Ross}\ \emph {et~al.}(2011)\citenamefont {Ross},
  \citenamefont {Savary}, \citenamefont {Gaulin},\ and\ \citenamefont
  {Balents}}]{ross_quantum_2011}%
  \BibitemOpen
  \bibfield  {author} {\bibinfo {author} {\bibfnamefont {K.~A.}\ \bibnamefont
  {Ross}}, \bibinfo {author} {\bibfnamefont {L.}~\bibnamefont {Savary}},
  \bibinfo {author} {\bibfnamefont {B.~D.}\ \bibnamefont {Gaulin}},\ and\
  \bibinfo {author} {\bibfnamefont {L.}~\bibnamefont {Balents}},\ }\bibfield
  {title} {\bibinfo {title} {Quantum {Excitations} in {Quantum} {Spin} {Ice}},\
  }\href {https://doi.org/10.1103/PhysRevX.1.021002} {\bibfield  {journal}
  {\bibinfo  {journal} {Phys. Rev. X}\ }\textbf {\bibinfo {volume} {1}},\
  \bibinfo {pages} {021002} (\bibinfo {year} {2011})}\BibitemShut {NoStop}%
\bibitem [{\citenamefont {Rau}\ and\ \citenamefont
  {Gingras}(2019)}]{Rau_Gingras_review_2019}%
  \BibitemOpen
  \bibfield  {author} {\bibinfo {author} {\bibfnamefont {J.~G.}\ \bibnamefont
  {Rau}}\ and\ \bibinfo {author} {\bibfnamefont {M.~J.}\ \bibnamefont
  {Gingras}},\ }\bibfield  {title} {\bibinfo {title} {Frustrated quantum
  rare-earth pyrochlores},\ }\href
  {https://doi.org/10.1146/annurev-conmatphys-022317-110520} {\bibfield
  {journal} {\bibinfo  {journal} {Annual Review of Condensed Matter Physics}\
  }\textbf {\bibinfo {volume} {10}},\ \bibinfo {pages} {357} (\bibinfo {year}
  {2019})},\ \Eprint
  {https://arxiv.org/abs/https://doi.org/10.1146/annurev-conmatphys-022317-110520}
  {https://doi.org/10.1146/annurev-conmatphys-022317-110520} \BibitemShut
  {NoStop}%
\bibitem [{\citenamefont {Derzhko}\ \emph {et~al.}(2020)\citenamefont
  {Derzhko}, \citenamefont {Hutak}, \citenamefont {Krokhmalskii}, \citenamefont
  {Schnack},\ and\ \citenamefont {Richter}}]{derzhko_adapting_2020}%
  \BibitemOpen
  \bibfield  {author} {\bibinfo {author} {\bibfnamefont {O.}~\bibnamefont
  {Derzhko}}, \bibinfo {author} {\bibfnamefont {T.}~\bibnamefont {Hutak}},
  \bibinfo {author} {\bibfnamefont {T.}~\bibnamefont {Krokhmalskii}}, \bibinfo
  {author} {\bibfnamefont {J.}~\bibnamefont {Schnack}},\ and\ \bibinfo {author}
  {\bibfnamefont {J.}~\bibnamefont {Richter}},\ }\bibfield  {title} {\bibinfo
  {title} {Adapting planck's route to investigate the thermodynamics of the
  spin-half pyrochlore heisenberg antiferromagnet},\ }\href
  {https://doi.org/10.1103/PhysRevB.101.174426} {\bibfield  {journal} {\bibinfo
   {journal} {Phys. Rev. B}\ }\textbf {\bibinfo {volume} {101}},\ \bibinfo
  {pages} {174426} (\bibinfo {year} {2020})}\BibitemShut {NoStop}%
\bibitem [{\citenamefont {Harris}\ \emph {et~al.}(1997)\citenamefont {Harris},
  \citenamefont {Bramwell}, \citenamefont {McMorrow}, \citenamefont {Zeiske},\
  and\ \citenamefont {Godfrey}}]{harris_geometrical_1997}%
  \BibitemOpen
  \bibfield  {author} {\bibinfo {author} {\bibfnamefont {M.~J.}\ \bibnamefont
  {Harris}}, \bibinfo {author} {\bibfnamefont {S.~T.}\ \bibnamefont
  {Bramwell}}, \bibinfo {author} {\bibfnamefont {D.~F.}\ \bibnamefont
  {McMorrow}}, \bibinfo {author} {\bibfnamefont {T.}~\bibnamefont {Zeiske}},\
  and\ \bibinfo {author} {\bibfnamefont {K.~W.}\ \bibnamefont {Godfrey}},\
  }\bibfield  {title} {\bibinfo {title} {Geometrical {Frustration} in the
  {Ferromagnetic} {Pyrochlore} {Ho$_2$Ti$_2$O$_7$}},\ }\href
  {https://doi.org/10.1103/PhysRevLett.79.2554} {\bibfield  {journal} {\bibinfo
   {journal} {Physical Review Letters}\ }\textbf {\bibinfo {volume} {79}},\
  \bibinfo {pages} {2554} (\bibinfo {year} {1997})}\BibitemShut {NoStop}%
\bibitem [{\citenamefont {Castelnovo}\ \emph {et~al.}(2012)\citenamefont
  {Castelnovo}, \citenamefont {Moessner},\ and\ \citenamefont
  {Sondhi}}]{castelnovo_spin_2012}%
  \BibitemOpen
  \bibfield  {author} {\bibinfo {author} {\bibfnamefont {C.}~\bibnamefont
  {Castelnovo}}, \bibinfo {author} {\bibfnamefont {R.}~\bibnamefont
  {Moessner}},\ and\ \bibinfo {author} {\bibfnamefont {S.}~\bibnamefont
  {Sondhi}},\ }\bibfield  {title} {\bibinfo {title} {Spin {Ice},
  {Fractionalization}, and {Topological} {Order}},\ }\href
  {https://doi.org/10.1146/annurev-conmatphys-020911-125058} {\bibfield
  {journal} {\bibinfo  {journal} {Annual Review of Condensed Matter Physics}\
  }\textbf {\bibinfo {volume} {3}},\ \bibinfo {pages} {35} (\bibinfo {year}
  {2012})}\BibitemShut {NoStop}%
\bibitem [{\citenamefont {Hagym\'asi}\ \emph {et~al.}(2021)\citenamefont
  {Hagym\'asi}, \citenamefont {Sch\"afer}, \citenamefont {Moessner},\ and\
  \citenamefont {Luitz}}]{hagymasi_prl_2021}%
  \BibitemOpen
  \bibfield  {author} {\bibinfo {author} {\bibfnamefont {I.}~\bibnamefont
  {Hagym\'asi}}, \bibinfo {author} {\bibfnamefont {R.}~\bibnamefont
  {Sch\"afer}}, \bibinfo {author} {\bibfnamefont {R.}~\bibnamefont
  {Moessner}},\ and\ \bibinfo {author} {\bibfnamefont {D.~J.}\ \bibnamefont
  {Luitz}},\ }\bibfield  {title} {\bibinfo {title} {Possible inversion symmetry
  breaking in the $s=1/2$ pyrochlore heisenberg magnet},\ }\href
  {https://doi.org/10.1103/PhysRevLett.126.117204} {\bibfield  {journal}
  {\bibinfo  {journal} {Phys. Rev. Lett.}\ }\textbf {\bibinfo {volume} {126}},\
  \bibinfo {pages} {117204} (\bibinfo {year} {2021})}\BibitemShut {NoStop}%
\bibitem [{\citenamefont {Astrakhantsev}\ \emph {et~al.}(2021)\citenamefont
  {Astrakhantsev}, \citenamefont {Westerhout}, \citenamefont {Tiwari},
  \citenamefont {Choo}, \citenamefont {Chen}, \citenamefont {Fischer},
  \citenamefont {Carleo},\ and\ \citenamefont
  {Neupert}}]{astrakhantsev_broken-symmetry_2021}%
  \BibitemOpen
  \bibfield  {author} {\bibinfo {author} {\bibfnamefont {N.}~\bibnamefont
  {Astrakhantsev}}, \bibinfo {author} {\bibfnamefont {T.}~\bibnamefont
  {Westerhout}}, \bibinfo {author} {\bibfnamefont {A.}~\bibnamefont {Tiwari}},
  \bibinfo {author} {\bibfnamefont {K.}~\bibnamefont {Choo}}, \bibinfo {author}
  {\bibfnamefont {A.}~\bibnamefont {Chen}}, \bibinfo {author} {\bibfnamefont
  {M.~H.}\ \bibnamefont {Fischer}}, \bibinfo {author} {\bibfnamefont
  {G.}~\bibnamefont {Carleo}},\ and\ \bibinfo {author} {\bibfnamefont
  {T.}~\bibnamefont {Neupert}},\ }\bibfield  {title} {\bibinfo {title}
  {Broken-symmetry ground states of the heisenberg model on the pyrochlore
  lattice},\ }\href {https://doi.org/10.1103/PhysRevX.11.041021} {\bibfield
  {journal} {\bibinfo  {journal} {Phys. Rev. X}\ }\textbf {\bibinfo {volume}
  {11}},\ \bibinfo {pages} {041021} (\bibinfo {year} {2021})}\BibitemShut
  {NoStop}%
\bibitem [{\citenamefont {Hering}\ \emph {et~al.}(2022)\citenamefont {Hering},
  \citenamefont {Noculak}, \citenamefont {Ferrari}, \citenamefont {Iqbal},\
  and\ \citenamefont {Reuther}}]{hering2021dimerization}%
  \BibitemOpen
  \bibfield  {author} {\bibinfo {author} {\bibfnamefont {M.}~\bibnamefont
  {Hering}}, \bibinfo {author} {\bibfnamefont {V.}~\bibnamefont {Noculak}},
  \bibinfo {author} {\bibfnamefont {F.}~\bibnamefont {Ferrari}}, \bibinfo
  {author} {\bibfnamefont {Y.}~\bibnamefont {Iqbal}},\ and\ \bibinfo {author}
  {\bibfnamefont {J.}~\bibnamefont {Reuther}},\ }\bibfield  {title} {\bibinfo
  {title} {Dimerization tendencies of the pyrochlore heisenberg
  antiferromagnet: A functional renormalization group perspective},\ }\href
  {https://doi.org/10.1103/PhysRevB.105.054426} {\bibfield  {journal} {\bibinfo
   {journal} {Phys. Rev. B}\ }\textbf {\bibinfo {volume} {105}},\ \bibinfo
  {pages} {054426} (\bibinfo {year} {2022})}\BibitemShut {NoStop}%
\bibitem [{\citenamefont {Sch\"afer}\ \emph {et~al.}(2022)\citenamefont
  {Sch\"afer}, \citenamefont {Placke}, \citenamefont {Benton},\ and\
  \citenamefont {Moessner}}]{schaefer2022abundance}%
  \BibitemOpen
  \bibfield  {author} {\bibinfo {author} {\bibfnamefont {R.}~\bibnamefont
  {Sch\"afer}}, \bibinfo {author} {\bibfnamefont {B.}~\bibnamefont {Placke}},
  \bibinfo {author} {\bibfnamefont {O.}~\bibnamefont {Benton}},\ and\ \bibinfo
  {author} {\bibfnamefont {R.}~\bibnamefont {Moessner}},\ }\href@noop {}
  {\bibinfo {title} {Abundance of hard-hexagon crystals in the quantum
  pyrochlore antiferromagnet}} (\bibinfo {year} {2022}),\ \Eprint
  {https://arxiv.org/abs/2210.07235} {arXiv:2210.07235 [cond-mat.str-el]}
  \BibitemShut {NoStop}%
\bibitem [{\citenamefont {Plumb}\ \emph {et~al.}(2019)\citenamefont {Plumb},
  \citenamefont {Changlani}, \citenamefont {Scheie}, \citenamefont {Zhang},
  \citenamefont {Krizan}, \citenamefont {Rodriguez-Rivera}, \citenamefont
  {Qiu}, \citenamefont {Winn}, \citenamefont {Cava},\ and\ \citenamefont
  {Broholm}}]{plumb_continuum_2019}%
  \BibitemOpen
  \bibfield  {author} {\bibinfo {author} {\bibfnamefont {K.~W.}\ \bibnamefont
  {Plumb}}, \bibinfo {author} {\bibfnamefont {H.~J.}\ \bibnamefont
  {Changlani}}, \bibinfo {author} {\bibfnamefont {A.}~\bibnamefont {Scheie}},
  \bibinfo {author} {\bibfnamefont {S.}~\bibnamefont {Zhang}}, \bibinfo
  {author} {\bibfnamefont {J.~W.}\ \bibnamefont {Krizan}}, \bibinfo {author}
  {\bibfnamefont {J.~A.}\ \bibnamefont {Rodriguez-Rivera}}, \bibinfo {author}
  {\bibfnamefont {Y.}~\bibnamefont {Qiu}}, \bibinfo {author} {\bibfnamefont
  {B.}~\bibnamefont {Winn}}, \bibinfo {author} {\bibfnamefont {R.~J.}\
  \bibnamefont {Cava}},\ and\ \bibinfo {author} {\bibfnamefont {C.~L.}\
  \bibnamefont {Broholm}},\ }\bibfield  {title} {\bibinfo {title} {Continuum of
  quantum fluctuations in a three-dimensional {S} = 1 {Heisenberg} magnet},\
  }\href {https://doi.org/10.1038/s41567-018-0317-3} {\bibfield  {journal}
  {\bibinfo  {journal} {Nature Physics}\ }\textbf {\bibinfo {volume} {15}},\
  \bibinfo {pages} {54} (\bibinfo {year} {2019})}\BibitemShut {NoStop}%
\bibitem [{\citenamefont {Niggemann}\ \emph {et~al.}(2021)\citenamefont
  {Niggemann}, \citenamefont {Sbierski},\ and\ \citenamefont
  {Reuther}}]{niggemann2021majorana}%
  \BibitemOpen
  \bibfield  {author} {\bibinfo {author} {\bibfnamefont {N.}~\bibnamefont
  {Niggemann}}, \bibinfo {author} {\bibfnamefont {B.}~\bibnamefont
  {Sbierski}},\ and\ \bibinfo {author} {\bibfnamefont {J.}~\bibnamefont
  {Reuther}},\ }\bibfield  {title} {\bibinfo {title} {Frustrated quantum spins
  at finite temperature: Pseudo-majorana functional renormalization group
  approach},\ }\href {https://doi.org/10.1103/PhysRevB.103.104431} {\bibfield
  {journal} {\bibinfo  {journal} {Phys. Rev. B}\ }\textbf {\bibinfo {volume}
  {103}},\ \bibinfo {pages} {104431} (\bibinfo {year} {2021})}\BibitemShut
  {NoStop}%
\bibitem [{\citenamefont {Niggemann}\ \emph
  {et~al.}(2022{\natexlab{a}})\citenamefont {Niggemann}, \citenamefont
  {Reuther},\ and\ \citenamefont {Sbierski}}]{niggemann2021quantitative}%
  \BibitemOpen
  \bibfield  {author} {\bibinfo {author} {\bibfnamefont {N.}~\bibnamefont
  {Niggemann}}, \bibinfo {author} {\bibfnamefont {J.}~\bibnamefont {Reuther}},\
  and\ \bibinfo {author} {\bibfnamefont {B.}~\bibnamefont {Sbierski}},\
  }\bibfield  {title} {\bibinfo {title} {{Quantitative functional
  renormalization for three-dimensional quantum Heisenberg models}},\ }\href
  {https://doi.org/10.21468/SciPostPhys.12.5.156} {\bibfield  {journal}
  {\bibinfo  {journal} {SciPost Phys.}\ }\textbf {\bibinfo {volume} {12}},\
  \bibinfo {pages} {156} (\bibinfo {year} {2022}{\natexlab{a}})}\BibitemShut
  {NoStop}%
\bibitem [{\citenamefont {White}(1992)}]{white_1992}%
  \BibitemOpen
  \bibfield  {author} {\bibinfo {author} {\bibfnamefont {S.~R.}\ \bibnamefont
  {White}},\ }\bibfield  {title} {\bibinfo {title} {Density matrix formulation
  for quantum renormalization groups},\ }\href
  {https://doi.org/10.1103/PhysRevLett.69.2863} {\bibfield  {journal} {\bibinfo
   {journal} {Phys. Rev. Lett.}\ }\textbf {\bibinfo {volume} {69}},\ \bibinfo
  {pages} {2863} (\bibinfo {year} {1992})}\BibitemShut {NoStop}%
\bibitem [{\citenamefont {White}(1993)}]{white_1993}%
  \BibitemOpen
  \bibfield  {author} {\bibinfo {author} {\bibfnamefont {S.~R.}\ \bibnamefont
  {White}},\ }\bibfield  {title} {\bibinfo {title} {Density-matrix algorithms
  for quantum renormalization groups},\ }\href
  {https://doi.org/10.1103/PhysRevB.48.10345} {\bibfield  {journal} {\bibinfo
  {journal} {Phys. Rev. B}\ }\textbf {\bibinfo {volume} {48}},\ \bibinfo
  {pages} {10345} (\bibinfo {year} {1993})}\BibitemShut {NoStop}%
\bibitem [{\citenamefont {Feiguin}\ and\ \citenamefont
  {White}(2005)}]{white_purification_2005}%
  \BibitemOpen
  \bibfield  {author} {\bibinfo {author} {\bibfnamefont {A.~E.}\ \bibnamefont
  {Feiguin}}\ and\ \bibinfo {author} {\bibfnamefont {S.~R.}\ \bibnamefont
  {White}},\ }\bibfield  {title} {\bibinfo {title} {Finite-temperature density
  matrix renormalization using an enlarged hilbert space},\ }\href
  {https://doi.org/10.1103/PhysRevB.72.220401} {\bibfield  {journal} {\bibinfo
  {journal} {Phys. Rev. B}\ }\textbf {\bibinfo {volume} {72}},\ \bibinfo
  {pages} {220401} (\bibinfo {year} {2005})}\BibitemShut {NoStop}%
\bibitem [{\citenamefont {Verstraete}\ \emph {et~al.}(2004)\citenamefont
  {Verstraete}, \citenamefont {Garc\'{\i}a-Ripoll},\ and\ \citenamefont
  {Cirac}}]{cirac_purification_2004}%
  \BibitemOpen
  \bibfield  {author} {\bibinfo {author} {\bibfnamefont {F.}~\bibnamefont
  {Verstraete}}, \bibinfo {author} {\bibfnamefont {J.~J.}\ \bibnamefont
  {Garc\'{\i}a-Ripoll}},\ and\ \bibinfo {author} {\bibfnamefont {J.~I.}\
  \bibnamefont {Cirac}},\ }\bibfield  {title} {\bibinfo {title} {Matrix product
  density operators: Simulation of finite-temperature and dissipative
  systems},\ }\href {https://doi.org/10.1103/PhysRevLett.93.207204} {\bibfield
  {journal} {\bibinfo  {journal} {Phys. Rev. Lett.}\ }\textbf {\bibinfo
  {volume} {93}},\ \bibinfo {pages} {207204} (\bibinfo {year}
  {2004})}\BibitemShut {NoStop}%
\bibitem [{\citenamefont {Schollw\"ock}(2005)}]{schollwock_review_2005}%
  \BibitemOpen
  \bibfield  {author} {\bibinfo {author} {\bibfnamefont {U.}~\bibnamefont
  {Schollw\"ock}},\ }\bibfield  {title} {\bibinfo {title} {The density-matrix
  renormalization group},\ }\href {https://doi.org/10.1103/RevModPhys.77.259}
  {\bibfield  {journal} {\bibinfo  {journal} {Rev. Mod. Phys.}\ }\textbf
  {\bibinfo {volume} {77}},\ \bibinfo {pages} {259} (\bibinfo {year}
  {2005})}\BibitemShut {NoStop}%
\bibitem [{\citenamefont {Schollw\"ock}(2011)}]{schollwock_review_2011}%
  \BibitemOpen
  \bibfield  {author} {\bibinfo {author} {\bibfnamefont {U.}~\bibnamefont
  {Schollw\"ock}},\ }\bibfield  {title} {\bibinfo {title} {The density-matrix
  renormalization group in the age of matrix product states},\ }\href
  {https://doi.org/https://doi.org/10.1016/j.aop.2010.09.012} {\bibfield
  {journal} {\bibinfo  {journal} {Annals of Physics}\ }\textbf {\bibinfo
  {volume} {326}},\ \bibinfo {pages} {96 } (\bibinfo {year} {2011})},\ \bibinfo
  {note} {january 2011 Special Issue}\BibitemShut {NoStop}%
\bibitem [{\citenamefont {Hallberg}(2006)}]{hallberg_review}%
  \BibitemOpen
  \bibfield  {author} {\bibinfo {author} {\bibfnamefont {K.~A.}\ \bibnamefont
  {Hallberg}},\ }\bibfield  {title} {\bibinfo {title} {New trends in density
  matrix renormalization},\ }\href {https://doi.org/10.1080/00018730600766432}
  {\bibfield  {journal} {\bibinfo  {journal} {Advances in Physics}\ }\textbf
  {\bibinfo {volume} {55}},\ \bibinfo {pages} {477} (\bibinfo {year} {2006})},\
  \Eprint {https://arxiv.org/abs/https://doi.org/10.1080/00018730600766432}
  {https://doi.org/10.1080/00018730600766432} \BibitemShut {NoStop}%
\bibitem [{\citenamefont {Niggemann}\ \emph
  {et~al.}(2022{\natexlab{b}})\citenamefont {Niggemann}, \citenamefont
  {Reuther},\ and\ \citenamefont {Sbierski}}]{Niggemann2021a}%
  \BibitemOpen
  \bibfield  {author} {\bibinfo {author} {\bibfnamefont {N.}~\bibnamefont
  {Niggemann}}, \bibinfo {author} {\bibfnamefont {J.}~\bibnamefont {Reuther}},\
  and\ \bibinfo {author} {\bibfnamefont {B.}~\bibnamefont {Sbierski}},\
  }\bibfield  {title} {\bibinfo {title} {{Quantitative functional
  renormalization for three-dimensional quantum Heisenberg models}},\ }\href
  {https://doi.org/10.21468/SciPostPhys.12.5.156} {\bibfield  {journal}
  {\bibinfo  {journal} {SciPost Phys.}\ }\textbf {\bibinfo {volume} {12}},\
  \bibinfo {pages} {156} (\bibinfo {year} {2022}{\natexlab{b}})}\BibitemShut
  {NoStop}%
\bibitem [{\citenamefont {Schaden}\ and\ \citenamefont
  {Reuther}(2023)}]{schaden2023bilinear}%
  \BibitemOpen
  \bibfield  {author} {\bibinfo {author} {\bibfnamefont {Y.}~\bibnamefont
  {Schaden}}\ and\ \bibinfo {author} {\bibfnamefont {J.}~\bibnamefont
  {Reuther}},\ }\bibfield  {title} {\bibinfo {title} {{Bilinear Majorana
  representations for spin operators with spin magnitudes $S > 1/2$}},\ }\href
  {https://doi.org/10.1103/PhysRevResearch.5.023067} {\bibfield  {journal}
  {\bibinfo  {journal} {Phys. Rev. Res.}\ }\textbf {\bibinfo {volume} {5}},\
  \bibinfo {pages} {023067} (\bibinfo {year} {2023})}\BibitemShut {NoStop}%
\bibitem [{\citenamefont {Iqbal}\ \emph {et~al.}(2017)\citenamefont {Iqbal},
  \citenamefont {M\"uller}, \citenamefont {Riedl}, \citenamefont {Reuther},
  \citenamefont {Rachel}, \citenamefont {Valent\'{\i}}, \citenamefont
  {Gingras}, \citenamefont {Thomale},\ and\ \citenamefont
  {Jeschke}}]{iqbal_pyrochlore_s05_2017}%
  \BibitemOpen
  \bibfield  {author} {\bibinfo {author} {\bibfnamefont {Y.}~\bibnamefont
  {Iqbal}}, \bibinfo {author} {\bibfnamefont {T.}~\bibnamefont {M\"uller}},
  \bibinfo {author} {\bibfnamefont {K.}~\bibnamefont {Riedl}}, \bibinfo
  {author} {\bibfnamefont {J.}~\bibnamefont {Reuther}}, \bibinfo {author}
  {\bibfnamefont {S.}~\bibnamefont {Rachel}}, \bibinfo {author} {\bibfnamefont
  {R.}~\bibnamefont {Valent\'{\i}}}, \bibinfo {author} {\bibfnamefont
  {M.~J.~P.}\ \bibnamefont {Gingras}}, \bibinfo {author} {\bibfnamefont
  {R.}~\bibnamefont {Thomale}},\ and\ \bibinfo {author} {\bibfnamefont {H.~O.}\
  \bibnamefont {Jeschke}},\ }\bibfield  {title} {\bibinfo {title} {Signatures
  of a gearwheel quantum spin liquid in a spin-$\frac{1}{2}$ pyrochlore
  molybdate heisenberg antiferromagnet},\ }\href
  {https://doi.org/10.1103/PhysRevMaterials.1.071201} {\bibfield  {journal}
  {\bibinfo  {journal} {Phys. Rev. Materials}\ }\textbf {\bibinfo {volume}
  {1}},\ \bibinfo {pages} {071201} (\bibinfo {year} {2017})}\BibitemShut
  {NoStop}%
\bibitem [{\citenamefont {Haegeman}\ \emph {et~al.}(2011)\citenamefont
  {Haegeman}, \citenamefont {Cirac}, \citenamefont {Osborne}, \citenamefont
  {Pi\ifmmode~\check{z}\else \v{z}\fi{}orn}, \citenamefont {Verschelde},\ and\
  \citenamefont {Verstraete}}]{haegeman_2011}%
  \BibitemOpen
  \bibfield  {author} {\bibinfo {author} {\bibfnamefont {J.}~\bibnamefont
  {Haegeman}}, \bibinfo {author} {\bibfnamefont {J.~I.}\ \bibnamefont {Cirac}},
  \bibinfo {author} {\bibfnamefont {T.~J.}\ \bibnamefont {Osborne}}, \bibinfo
  {author} {\bibfnamefont {I.}~\bibnamefont {Pi\ifmmode~\check{z}\else
  \v{z}\fi{}orn}}, \bibinfo {author} {\bibfnamefont {H.}~\bibnamefont
  {Verschelde}},\ and\ \bibinfo {author} {\bibfnamefont {F.}~\bibnamefont
  {Verstraete}},\ }\bibfield  {title} {\bibinfo {title} {Time-dependent
  variational principle for quantum lattices},\ }\href
  {https://doi.org/10.1103/PhysRevLett.107.070601} {\bibfield  {journal}
  {\bibinfo  {journal} {Phys. Rev. Lett.}\ }\textbf {\bibinfo {volume} {107}},\
  \bibinfo {pages} {070601} (\bibinfo {year} {2011})}\BibitemShut {NoStop}%
\bibitem [{\citenamefont {Haegeman}\ \emph {et~al.}(2016)\citenamefont
  {Haegeman}, \citenamefont {Lubich}, \citenamefont {Oseledets}, \citenamefont
  {Vandereycken},\ and\ \citenamefont {Verstraete}}]{haegeman_2016}%
  \BibitemOpen
  \bibfield  {author} {\bibinfo {author} {\bibfnamefont {J.}~\bibnamefont
  {Haegeman}}, \bibinfo {author} {\bibfnamefont {C.}~\bibnamefont {Lubich}},
  \bibinfo {author} {\bibfnamefont {I.}~\bibnamefont {Oseledets}}, \bibinfo
  {author} {\bibfnamefont {B.}~\bibnamefont {Vandereycken}},\ and\ \bibinfo
  {author} {\bibfnamefont {F.}~\bibnamefont {Verstraete}},\ }\bibfield  {title}
  {\bibinfo {title} {Unifying time evolution and optimization with matrix
  product states},\ }\href {https://doi.org/10.1103/PhysRevB.94.165116}
  {\bibfield  {journal} {\bibinfo  {journal} {Phys. Rev. B}\ }\textbf {\bibinfo
  {volume} {94}},\ \bibinfo {pages} {165116} (\bibinfo {year}
  {2016})}\BibitemShut {NoStop}%
\bibitem [{\citenamefont {Schäfer}\ \emph {et~al.}(2020)\citenamefont
  {Schäfer}, \citenamefont {Hagymási}, \citenamefont {Moessner},\ and\
  \citenamefont {Luitz}}]{schafer_pyrochlore_2020}%
  \BibitemOpen
  \bibfield  {author} {\bibinfo {author} {\bibfnamefont {R.}~\bibnamefont
  {Schäfer}}, \bibinfo {author} {\bibfnamefont {I.}~\bibnamefont {Hagymási}},
  \bibinfo {author} {\bibfnamefont {R.}~\bibnamefont {Moessner}},\ and\
  \bibinfo {author} {\bibfnamefont {D.~J.}\ \bibnamefont {Luitz}},\ }\bibfield
  {title} {\bibinfo {title} {Pyrochlore $s=\frac{1}{2}$ {Heisenberg}
  antiferromagnet at finite temperature},\ }\href
  {https://doi.org/10.1103/PhysRevB.102.054408} {\bibfield  {journal} {\bibinfo
   {journal} {Physical Review B}\ }\textbf {\bibinfo {volume} {102}},\ \bibinfo
  {pages} {054408} (\bibinfo {year} {2020})},\ \bibinfo {note} {publisher:
  American Physical Society}\BibitemShut {NoStop}%
\bibitem [{\citenamefont {Paeckel}\ \emph {et~al.}(2019)\citenamefont
  {Paeckel}, \citenamefont {Köhler}, \citenamefont {Swoboda}, \citenamefont
  {Manmana}, \citenamefont {Schollw\"ock},\ and\ \citenamefont
  {Hubig}}]{hubig_review_2019}%
  \BibitemOpen
  \bibfield  {author} {\bibinfo {author} {\bibfnamefont {S.}~\bibnamefont
  {Paeckel}}, \bibinfo {author} {\bibfnamefont {T.}~\bibnamefont {Köhler}},
  \bibinfo {author} {\bibfnamefont {A.}~\bibnamefont {Swoboda}}, \bibinfo
  {author} {\bibfnamefont {S.~R.}\ \bibnamefont {Manmana}}, \bibinfo {author}
  {\bibfnamefont {U.}~\bibnamefont {Schollw\"ock}},\ and\ \bibinfo {author}
  {\bibfnamefont {C.}~\bibnamefont {Hubig}},\ }\bibfield  {title} {\bibinfo
  {title} {Time-evolution methods for matrix-product states},\ }\href
  {https://doi.org/https://doi.org/10.1016/j.aop.2019.167998} {\bibfield
  {journal} {\bibinfo  {journal} {Annals of Physics}\ }\textbf {\bibinfo
  {volume} {411}},\ \bibinfo {pages} {167998} (\bibinfo {year}
  {2019})}\BibitemShut {NoStop}%
\bibitem [{\citenamefont {Hubig}\ \emph {et~al.}(2015)\citenamefont {Hubig},
  \citenamefont {McCulloch}, \citenamefont {Schollw\"ock},\ and\ \citenamefont
  {Wolf}}]{hubig_2015}%
  \BibitemOpen
  \bibfield  {author} {\bibinfo {author} {\bibfnamefont {C.}~\bibnamefont
  {Hubig}}, \bibinfo {author} {\bibfnamefont {I.~P.}\ \bibnamefont
  {McCulloch}}, \bibinfo {author} {\bibfnamefont {U.}~\bibnamefont
  {Schollw\"ock}},\ and\ \bibinfo {author} {\bibfnamefont {F.~A.}\ \bibnamefont
  {Wolf}},\ }\bibfield  {title} {\bibinfo {title} {Strictly single-site dmrg
  algorithm with subspace expansion},\ }\href
  {https://doi.org/10.1103/PhysRevB.91.155115} {\bibfield  {journal} {\bibinfo
  {journal} {Phys. Rev. B}\ }\textbf {\bibinfo {volume} {91}},\ \bibinfo
  {pages} {155115} (\bibinfo {year} {2015})}\BibitemShut {NoStop}%
\bibitem [{\citenamefont {Hubig}(2017)}]{hubig17:_symmet_protec_tensor_networ}%
  \BibitemOpen
  \bibfield  {author} {\bibinfo {author} {\bibfnamefont {C.}~\bibnamefont
  {Hubig}},\ }\emph {\bibinfo {title} {Symmetry-Protected Tensor Networks}},\
  \href {https://edoc.ub.uni-muenchen.de/21348/} {Ph.D. thesis},\ \bibinfo
  {school} {LMU M\"unchen} (\bibinfo {year} {2017})\BibitemShut {NoStop}%
\bibitem [{\citenamefont {Hagym\'asi}\ \emph {et~al.}(2022)\citenamefont
  {Hagym\'asi}, \citenamefont {Noculak},\ and\ \citenamefont
  {Reuther}}]{hagymasi2022prb}%
  \BibitemOpen
  \bibfield  {author} {\bibinfo {author} {\bibfnamefont {I.}~\bibnamefont
  {Hagym\'asi}}, \bibinfo {author} {\bibfnamefont {V.}~\bibnamefont
  {Noculak}},\ and\ \bibinfo {author} {\bibfnamefont {J.}~\bibnamefont
  {Reuther}},\ }\bibfield  {title} {\bibinfo {title} {Enhanced
  symmetry-breaking tendencies in the $s=1$ pyrochlore antiferromagnet},\
  }\href {https://doi.org/10.1103/PhysRevB.106.235137} {\bibfield  {journal}
  {\bibinfo  {journal} {Phys. Rev. B}\ }\textbf {\bibinfo {volume} {106}},\
  \bibinfo {pages} {235137} (\bibinfo {year} {2022})}\BibitemShut {NoStop}%
\bibitem [{\citenamefont {Hubig}\ \emph {et~al.}(2018)\citenamefont {Hubig},
  \citenamefont {Haegeman},\ and\ \citenamefont
  {Schollw\"ock}}]{hubig_prb_2018}%
  \BibitemOpen
  \bibfield  {author} {\bibinfo {author} {\bibfnamefont {C.}~\bibnamefont
  {Hubig}}, \bibinfo {author} {\bibfnamefont {J.}~\bibnamefont {Haegeman}},\
  and\ \bibinfo {author} {\bibfnamefont {U.}~\bibnamefont {Schollw\"ock}},\
  }\bibfield  {title} {\bibinfo {title} {Error estimates for extrapolations
  with matrix-product states},\ }\href
  {https://doi.org/10.1103/PhysRevB.97.045125} {\bibfield  {journal} {\bibinfo
  {journal} {Phys. Rev. B}\ }\textbf {\bibinfo {volume} {97}},\ \bibinfo
  {pages} {045125} (\bibinfo {year} {2018})}\BibitemShut {NoStop}%
\bibitem [{\citenamefont {Schneider}\ \emph {et~al.}(2023)\citenamefont
  {Schneider}, \citenamefont {Reuther}, \citenamefont {Gonzalez}, \citenamefont
  {Sbierski},\ and\ \citenamefont {Niggemann}}]{schneider2023temperature}%
  \BibitemOpen
  \bibfield  {author} {\bibinfo {author} {\bibfnamefont {B.}~\bibnamefont
  {Schneider}}, \bibinfo {author} {\bibfnamefont {J.}~\bibnamefont {Reuther}},
  \bibinfo {author} {\bibfnamefont {M.~G.}\ \bibnamefont {Gonzalez}}, \bibinfo
  {author} {\bibfnamefont {B.}~\bibnamefont {Sbierski}},\ and\ \bibinfo
  {author} {\bibfnamefont {N.}~\bibnamefont {Niggemann}},\ }\href@noop {}
  {\bibinfo {title} {Temperature flow in pseudo-majorana functional
  renormalization for quantum spins}} (\bibinfo {year} {2023}),\ \Eprint
  {https://arxiv.org/abs/2312.14838} {arXiv:2312.14838 [cond-mat.str-el]}
  \BibitemShut {NoStop}%
\bibitem [{\citenamefont {Müller}\ \emph
  {et~al.}(2023{\natexlab{a}})\citenamefont {Müller}, \citenamefont {Kiese},
  \citenamefont {Niggemann}, \citenamefont {Sbierski}, \citenamefont {Reuther},
  \citenamefont {Trebst}, \citenamefont {Thomale},\ and\ \citenamefont
  {Iqbal}}]{muller2023pseudo}%
  \BibitemOpen
  \bibfield  {author} {\bibinfo {author} {\bibfnamefont {T.}~\bibnamefont
  {Müller}}, \bibinfo {author} {\bibfnamefont {D.}~\bibnamefont {Kiese}},
  \bibinfo {author} {\bibfnamefont {N.}~\bibnamefont {Niggemann}}, \bibinfo
  {author} {\bibfnamefont {B.}~\bibnamefont {Sbierski}}, \bibinfo {author}
  {\bibfnamefont {J.}~\bibnamefont {Reuther}}, \bibinfo {author} {\bibfnamefont
  {S.}~\bibnamefont {Trebst}}, \bibinfo {author} {\bibfnamefont
  {R.}~\bibnamefont {Thomale}},\ and\ \bibinfo {author} {\bibfnamefont
  {Y.}~\bibnamefont {Iqbal}},\ }\href@noop {} {\bibinfo {title} {Pseudo-fermion
  functional renormalization group for spin models}} (\bibinfo {year}
  {2023}{\natexlab{a}}),\ \Eprint {https://arxiv.org/abs/2307.10359}
  {arXiv:2307.10359 [cond-mat.str-el]} \BibitemShut {NoStop}%
\bibitem [{\citenamefont {Baez}\ and\ \citenamefont {Reuther}(2017)}]{Baez17}%
  \BibitemOpen
  \bibfield  {author} {\bibinfo {author} {\bibfnamefont {M.~L.}\ \bibnamefont
  {Baez}}\ and\ \bibinfo {author} {\bibfnamefont {J.}~\bibnamefont {Reuther}},\
  }\bibfield  {title} {\bibinfo {title} {{Numerical treatment of spin systems
  with unrestricted spin length $S$: A functional renormalization group
  study}},\ }\href {https://doi.org/10.1103/PhysRevB.96.045144} {\bibfield
  {journal} {\bibinfo  {journal} {Phys. Rev. B}\ }\textbf {\bibinfo {volume}
  {96}},\ \bibinfo {pages} {045144} (\bibinfo {year} {2017})}\BibitemShut
  {NoStop}%
\bibitem [{\citenamefont {Sandvik}(2010)}]{sandvik_computational_2010}%
  \BibitemOpen
  \bibfield  {author} {\bibinfo {author} {\bibfnamefont {A.~W.}\ \bibnamefont
  {Sandvik}},\ }\bibfield  {title} {\bibinfo {title} {Computational {Studies}
  of {Quantum} {Spin} {Systems}},\ }\href {https://doi.org/10.1063/1.3518900}
  {\bibfield  {journal} {\bibinfo  {journal} {AIP Conference Proceedings}\
  }\textbf {\bibinfo {volume} {1297}},\ \bibinfo {pages} {135} (\bibinfo {year}
  {2010})}\BibitemShut {NoStop}%
\bibitem [{\citenamefont {Müller}\ \emph
  {et~al.}(2023{\natexlab{b}})\citenamefont {Müller}, \citenamefont {Kiese},
  \citenamefont {Niggemann}, \citenamefont {Sbierski}, \citenamefont {Reuther},
  \citenamefont {Trebst}, \citenamefont {Thomale},\ and\ \citenamefont
  {Iqbal}}]{mueller2023pseudofermion}%
  \BibitemOpen
  \bibfield  {author} {\bibinfo {author} {\bibfnamefont {T.}~\bibnamefont
  {Müller}}, \bibinfo {author} {\bibfnamefont {D.}~\bibnamefont {Kiese}},
  \bibinfo {author} {\bibfnamefont {N.}~\bibnamefont {Niggemann}}, \bibinfo
  {author} {\bibfnamefont {B.}~\bibnamefont {Sbierski}}, \bibinfo {author}
  {\bibfnamefont {J.}~\bibnamefont {Reuther}}, \bibinfo {author} {\bibfnamefont
  {S.}~\bibnamefont {Trebst}}, \bibinfo {author} {\bibfnamefont
  {R.}~\bibnamefont {Thomale}},\ and\ \bibinfo {author} {\bibfnamefont
  {Y.}~\bibnamefont {Iqbal}},\ }\href@noop {} {\bibinfo {title} {Pseudo-fermion
  functional renormalization group for spin models}} (\bibinfo {year}
  {2023}{\natexlab{b}}),\ \Eprint {https://arxiv.org/abs/2307.10359}
  {arXiv:2307.10359 [cond-mat.str-el]} \BibitemShut {NoStop}%
\bibitem [{\citenamefont {Schneider}\ \emph {et~al.}(2022)\citenamefont
  {Schneider}, \citenamefont {Kiese},\ and\ \citenamefont
  {Sbierski}}]{schneider2022_taming}%
  \BibitemOpen
  \bibfield  {author} {\bibinfo {author} {\bibfnamefont {B.}~\bibnamefont
  {Schneider}}, \bibinfo {author} {\bibfnamefont {D.}~\bibnamefont {Kiese}},\
  and\ \bibinfo {author} {\bibfnamefont {B.}~\bibnamefont {Sbierski}},\
  }\bibfield  {title} {\bibinfo {title} {Taming pseudofermion functional
  renormalization for quantum spins: Finite temperatures and the popov-fedotov
  trick},\ }\href {https://doi.org/10.1103/PhysRevB.106.235113} {\bibfield
  {journal} {\bibinfo  {journal} {Phys. Rev. B}\ }\textbf {\bibinfo {volume}
  {106}},\ \bibinfo {pages} {235113} (\bibinfo {year} {2022})}\BibitemShut
  {NoStop}%
\bibitem [{\citenamefont {Hubig}\ \emph {et~al.}()\citenamefont {Hubig},
  \citenamefont {Lachenmaier}, \citenamefont {Linden}, \citenamefont
  {Reinhard}, \citenamefont {Stenzel}, \citenamefont {Swoboda},\ and\
  \citenamefont {Grundner}}]{hubig:_syten_toolk}%
  \BibitemOpen
  \bibfield  {author} {\bibinfo {author} {\bibfnamefont {C.}~\bibnamefont
  {Hubig}}, \bibinfo {author} {\bibfnamefont {F.}~\bibnamefont {Lachenmaier}},
  \bibinfo {author} {\bibfnamefont {N.-O.}\ \bibnamefont {Linden}}, \bibinfo
  {author} {\bibfnamefont {T.}~\bibnamefont {Reinhard}}, \bibinfo {author}
  {\bibfnamefont {L.}~\bibnamefont {Stenzel}}, \bibinfo {author} {\bibfnamefont
  {A.}~\bibnamefont {Swoboda}},\ and\ \bibinfo {author} {\bibfnamefont
  {M.}~\bibnamefont {Grundner}},\ }\href {https://syten.eu} {\bibinfo {title}
  {The \textsc{SyTen} toolkit}}\BibitemShut {NoStop}%
\bibitem [{Note1()}]{Note1}%
  \BibitemOpen
  \bibinfo {note} {This must only be implemented once for an arbitrary lattice.
  An examplary code implementation can be found in our publicly available
  package \protect \href
  {https://github.com/NilsNiggemann/SpinFRGLattices.jl/blob/dac0bc73ec6e28f82e9c445d8a5a78671ccb48b9/src/SpinSGeneralization.jl}{SpinFRGLattices.jl},
  where a given lattice geometry may simply be modified to obtain the
  corresponding effective spin-$S$ model.}\BibitemShut {Stop}%
\end{thebibliography}%
\end{document}